\documentclass[11pt]{amsart}
\usepackage{hyperref}
\usepackage{amssymb}
\usepackage{epsfig}
\usepackage{comment}
\usepackage{amsmath}
\usepackage{subcaption}
\usepackage[section]{placeins}
\usepackage{mathrsfs}
\usepackage{graphicx}
\usepackage[notrig]{physics}
\usepackage{units}
\usepackage{xcolor}
\usepackage{algorithm,algorithmic}
\usepackage{blindtext}
\usepackage{geometry}
 \graphicspath{ {./Figures/} }

\numberwithin{equation}{section}

\newtheorem{thm}{Theorem}[section]
\newtheorem{prop}[thm]{Proposition}

\theoremstyle{definition}

\newtheorem{example}[thm]{Example}

\theoremstyle{remark}

\begin{document}

\title[Data-Driven Molecular Dynamics]{Interatomic-Potential-Free, Data-Driven Molecular Dynamics}

\author[J.~Bulin, J.~Hamaekers, M.~P.~Ariza and M.~Ortiz]{
J.~Bulin${}^1$, J.~Hamaekers${}^1$, M.~P.~Ariza${}^2$ and M.~Ortiz${}^{3,4}$
}

\address
{
    ${}^1$Fraunhofer Institute for Algorithms and Scientific Computing, Schloss Birlinghoven, 53757 Sankt Augustin, Germany \\ \newline\indent
    ${}^2$Escuela T{\'e}cnica Superior de Ingenier{\'i}a, Universidad de Sevilla, Sevilla 41092, Spain \\ \newline\indent
    ${}^3$Hausdorff Center for Mathematics, Universit\"at Bonn, Endenicher Allee 60, 53115 Bonn, Germany \\ \newline\indent
    ${}^4$Division of Engineering and Applied Science, California Institute of Technology, 1200 E.~California Blvd., Pasadena, CA 91125.
}

%
%

\begin{abstract}
We present a Data-Driven (DD) paradigm that enables molecular dynamics calculations to be performed directly from sampled force-field data such as obtained, e.~g., from {\sl ab initio} calculations, thereby eschewing the conventional step of modeling the data by empirical interatomic potentials entirely. The data required by the DD solvers consists of local atomic configurations and corresponding atomic forces and is, therefore, {\sl fundamental}, i.~e., it is not beholden to any particular model. The resulting DD solvers, including a fully explicit DD-Verlet algorithm, are provably convergent and exhibit robust convergence with respect to the data in selected test cases. We present an example of application to ${\sl C}_{60}$ buckminsterfullerenes that showcases the feasibility, range and scope of the DD molecular dynamics paradigm.
\end{abstract}



\maketitle




\section{Introduction}


Present-day computational chemistry methodology, including all-electron calculations, Density Functional Theory (DFT), and other {\sl ab initio} paradigms, enables the generation of vast sets of parameter-free force-field data for complex molecular systems (cf., e.~g., \cite{Xu:2018} and references therein). However, for large systems of atoms molecular dynamics based on empirical interatomic potentials remains the only method of choice. In that setting, the empirical potentials required in the calculations are, inevitably, a main source of empiricism and error. Indeed, at present there does not exist a rigorous theoretical means of generating sequences $V_h$ of approximate interatomic potentials that are guaranteed to converge to the underlying--and unknown--exact potential $V_{\sl ab\ initio}$ in a manner that additionally ensures convergence of trajectories. Instead, empirical potentials based on {\sl ad hoc} assumptions and parametrizations are contrived and fitted to data without global control of errors or convergence guarantees (cf., e.~g., \cite{KIM} for the state of the art of empirical interatomic potentials).

The availability of vast quantities of high-fidelity force-field data (cf., e.~g., \cite{NoMaD, MP, NIST}) suggests a game-changing paradigm shift whereby the data themselves are the sole basis of molecular dynamics calculations and the conventional interatomic-potential modeling step is eschewed altogether. Evidently, such a strict Data-Driven (DD) paradigm, if feasible, would have the immediate beneficial consequences of eliminating the biases, loss of information, empiricism and error that inevitably afflict empirical interatomic potentials.

In this work we present a provably convergent strict DD paradigm for molecular dynamics based solely on force-field data and demonstrate the feasibility of the paradigm with the aid of selected examples. The resulting DD solvers return approximate trajectories that converge to those of the underlying--and unknown--exact potential when computed from force-field data sets of increasing fidelity. Such data sets can be generated {\sl ab initio}, stored, reused, merged and adapted to specific application domains along the tenets of {\sl active learning} \cite{Settles:2012}.

The specific DD paradigm considered in this work builds on similar approaches for static problems \cite{Kirch:2016, Kirchdoerfer:2017} and dynamic problems \cite{Kirchdoerfer:2018}. Specifically, the trajectories of the system are thought to take place in a {\sl phase space} of atomic configurations and force fields, cf.~Section~\ref{N3q49y}. The objective is then to determine trajectories that satisfy Newton's laws of motion exactly, viewed as a constraint on forces and accelerations, while remaining as close as possible to a given force-field data set in the sense of some suitable distance, cf.~Section \ref{2bzQRE}. We show that the resulting problem for the trajectories can be given the structure of an optimal control problem, cf.~Section \ref{7Iw6wu}, or a game-theoretical problem, cf.~Section \ref{4iH7kb}. In this latter formulation, we also show that the DD problem can be expressed in terms of an effective, or {\sl learned}, force field. However, the effective force field is only implied and need not be computed explicitly in calculations.

We note, cf.~Section \ref{ZJ1cqZ}, that a natural distance between local clusters of atoms that is invariant under relabeling of the atoms is provided by the Wasserstein distance of optimal transport \cite{Evans:1997}. In addition, the force-field data is invariant under the action of the Euclidean group of translations and orthogonal transformation, which requires the evaluation of distances between entire orbits of the Euclidean group. The calculation of distances between local atomic configurations is one of the main computational bottlenecks of the DD solver and, for large atomic clusters, requires the use of relaxation techniques {\sl \`a la} Kantorovich \cite{Evans:1997} and possibly interior-point regularizations such as max-ent \cite{ShuCherng:2001}. In addition, the DD solver entails frequent searches in large force-field data sets. Here again, efficient search algorithms originally developed for Big Data applications are in existence (cf., e.~g., \cite{Eggersmann:2021} and references therein) and can be deployed as part of the DD solver.

The convergence of the DD solvers with respect to the data, including a fully-explicit DD-Verlet algorithm derived by time discretization, cf.~Sections \ref{He0WpI} and \ref{D63kfb}, can be verified mathematically under simple data sampling scenarios, cf.~Section \ref{qQV1fx}, or numerically based on selected test cases, cf.~Section \ref{49azWB}. As a proof of concept, we also present a simple application of the DD-Verlet solver to ${\rm C}_{60}$ buckminsterfullerenes based on synthetic data sampled from the Stillinger-Weber potential \cite{SW:1985}, cf.~Section \ref{X1H1xV}. In all these cases, the ability of the DD solvers to compute qualitatively correct trajectories from relatively small data sets is remarkable.

\section{Classical molecular dynamics}

In this section, we define the class of problems under consideration and set forth notational conventions. We begin by considering classical dynamics without any assumptions regarding the structure and properties of the force field such as locality, Euclidean invariance or other symmetries. The consequences of such additional structures are elucidated in subsequent sections. We also concern ourselves with the reformulation of classical dynamics as a variational problem for trajectories in phase space, which sets forth a natural framework for the Data-Driven paradigm developed subsequently.

\subsection{General interatomic potentials}

We consider a system of $N$ atoms adopting configurations described by coordinates $r_i = (r_i^\alpha)_{\alpha=1}^3$, $i=1,\dots,N$, in three-dimensional Euclidean space, which we identify with $\mathbb{R}^3$ equipped with the standard Euclidean metric. For shorthand, we denote by $r \equiv \{r_i\}_{i=1}^N \in \mathbb{R}^{3N}$ the collection of all the position vectors of the atoms in the system and refer to $\mathbb{R}^{3N}$ as the {\sl configuration space}.

The motion of the atoms obeys Newton's second law
\begin{subequations}\label{5xvxXr}
\begin{align}
    & \label{0kzJ6C}
    m_i \ddot{r}_i(t) + f_i(r(t)) = f_i^{\rm ext}(t) ,
    \\ & \label{vKAl5W}
    r(0) = r_0,
    \quad
    \dot{r}(0) = v_0 ,
\end{align}
\end{subequations}
where $t$ denotes time, $m_i$ is the mass of atom $i$, $f_i(r)$ is the force on atom $i$ due to atomic interactions, $f_i^{\rm ext}(t)$ are applied forces, $r_0 \in \mathbb{R}^{3N}$ is the initial configuration, $v_0 \in \mathbb{R}^{3N}$ is the initial velocity field and a superimposed dot denotes time differentiation. The {\sl force field} of the system is the function $f(r) \equiv \{f_i(r)\}_{i=1}^N$ from configuration space $\mathbb{R}^{3N}$ to {\sl force-field space} $\mathbb{R}^{3N}$.

We note that the configuration space, identified with $\mathbb{R}^{3N}$, is to be regarded as an affine space of points, whereas the force-field space, while also identified with $\mathbb{R}^{3N}$, is to be regarded as a vector space. This distinction is consequential when metrizing configuration space, where a distance between sets of points must be defined, cf.~Section~\ref{ZJ1cqZ}\footnote{More general formulations allow for the configuration space to be a smooth manifold with force fields taking values in the corresponding cotangent spaces \cite{AMR:1988}, but that degree of generality is not required here.}.

The interaction forces are {\sl conservative} if there exists an interatomic potential $V(r)$ such that
\begin{equation}\label{1UpoOs}
    f_i(r) = \frac{\partial V}{\partial r_i}(r) .
\end{equation}
Examples of classical interatomic potentials commonly used in practice are presented, e.~g., in \cite{Finnis:2003}. Recent work aimed at formulating a framework for developing physics-based and machine learning interatomic potentials is exemplified by \cite{WEN2022108218}.

\subsection{Phase-space reformulation}\label{N3q49y}

An alternative set-oriented representation of the force field $f(r)$ is to view it as a graph $D$ in {\sl phase space} $Z = \mathbb{R}^{3N} \times \mathbb{R}^{3N}$, namely, the space of all pairs $(r,f)$ of system configurations and forces\footnote{It should be noted that the term {\sl phase space} in classical dynamics is often applied to the space of positions and momenta, whereas in this work we use the term to signify the space of positions and forces.}. In this representation, the force field is regarded as a material-specific $3N$-dimensional manifold, or graph, in $6N$-dimensional phase space $Z$ characterizing the entire range of possible atomic interactions.

We shall additionally denote by $\mathcal{Z}$ the linear space of all trial trajectories $z(\cdot) \equiv (r(\cdot),f(\cdot))$ of the system over a given time interval $[0,T]$. Thus, the elements of $\mathcal{Z}$ are curves $z(\cdot)$ in $Z$ parameterized by time, not necessarily satisfying the equations of motion or compatible with the force field of the system.

{\sl NB: In order to carefully differentiate between points and trajectories, henceforth we shall denote by $r$, $f$ and $z=(r,f)$ points in configuration space $\mathbb{R}^{3N}$, force space $\mathbb{R}^{3N}$ and phase space $Z = \mathbb{R}^{3N} \times \mathbb{R}^{3N}$, respectively; and we shall denote by $r(\cdot)$, $f(\cdot)$ and $z(\cdot)=(r(\cdot),f(\cdot))$ trajectories in the same spaces defined over a given time interval $[0,T]$. In particular, $r(t)$, $f(t)$ and $z(t)=(r(t),f(t))$ denote the values of trajectories $r(\cdot)$, $f(\cdot)$ and $z(\cdot)=(r(\cdot),f(\cdot))$  at time $t \in [0,T]$, respectively.}

For a specific material, the phase-space trajectories of the system must take values in the force-field graph $D$ at all times, i.~e., they must be contained in the set
\begin{equation}
    \mathcal{D}
    =
    \big\{
        (r(\cdot),f(\cdot)) \in \mathcal{Z} \, : \,
        (r(t),f(t)) \in D, \; t \in [0,T]
    \big\} .
\end{equation}
Thus, $\mathcal{D}$ is the set of phase-space trajectories compatible with the force field of the system and may thus be regarded as a material-specific trajectory set, or {\sl material set} for short. Euclidean invariance requires that $(r,f) \in D$ if and only if $(Q \,  (r - c), {Q} f) \in D$ for all translations $c \in \mathbb{R}^3$ and orthogonal transformations $Q \in O(3)$. Thus, $D$ must contain entire $E(3)$-{\sl orbits} and be invariant under the action of the {\sl Euclidean group} $E(3)$. In addition, we recall that the configuration space consists of point sets and should, therefore, be invariant under relabeling of the atoms, i.~e., $(r_i,f_i)_{i=1}^N \in D$ if and only if $(r_{\sigma(i)},f_{\sigma(i)})_{i=1}^N \in D$ for all permutations $\sigma$ of the index set $\{1,\dots,N\}$.

The physically admissible trajectories of the system are additionally subject to the constraint set forth by the equations of motion (\ref{5xvxXr}). The collection of all such admissible trajectories defines the admissible-trajectory set, or {\sl admissible set} for short,
\begin{equation}\label{T2yCfR}
\begin{split}
    &
    \mathcal{E}
    =
    \big\{
        z(\cdot) \equiv (r(\cdot),f(\cdot)) \in \mathcal{Z}
        \, : \, \\ & \qquad\qquad
        m_i \ddot{r}_i(t) + f_i(t) = f_i^{\rm ext}(t),
        \ t \in [0,T] ;
        \ r(0) = r_0,\ \dot{r}(0) = v_0
    \big\} .
\end{split}
\end{equation}
We note that the admissible set $\mathcal{E}$ depends parametrically on the applied force field $f^{\rm ext}(t)$ and the initial conditions $(r_0,v_0)$. However, the trial trajectories $(r(\cdot),f(\cdot))$  in $\mathcal{E}$  need not be solutions of the initial-value problem (\ref{5xvxXr}), since we do not require $r(t)$ and $f(t)$ to be related by the force field $f(r)$ of the material.

Evidently, the actual phase-space trajectories $(r(\cdot),f(\cdot))$ of the system, if they exist, lie in the intersection $\mathcal{D} \, \cap \, \mathcal{E}$, i.~e., are the admissible trajectories that are consistent with the force field of the material, or, equivalently, the material trajectories that are consistent with the equations of motion and initial conditions.

\section{Data-Driven reformulation}\label{2bzQRE}

Suppose that, as is often the case in practice, the graph $D$ of the force field is not known in its entirety, but only through an approximating sequence of data sets $D_h$, $h=0,1,\dots$. For instance, the sequence $(D_h)$ may consist of increasing collections of points $(r,f)$ in phase space $Z$ obtained by means of ancillary {\sl ab initio} calculations or by some other means. As noted before, $D_h$ must contain entire $E(3)$-orbits in order to be invariant under the action of the Euclidean group $E(3)$. In addition, the configurations sampled in $D_h$ are clusters of points and, therefore, their representation must be invariant under relabeling of the atoms.

\begin{figure}[ht]
\begin{center}
	\begin{subfigure}{0.45\textwidth}\caption{} \includegraphics[width=0.99\linewidth]{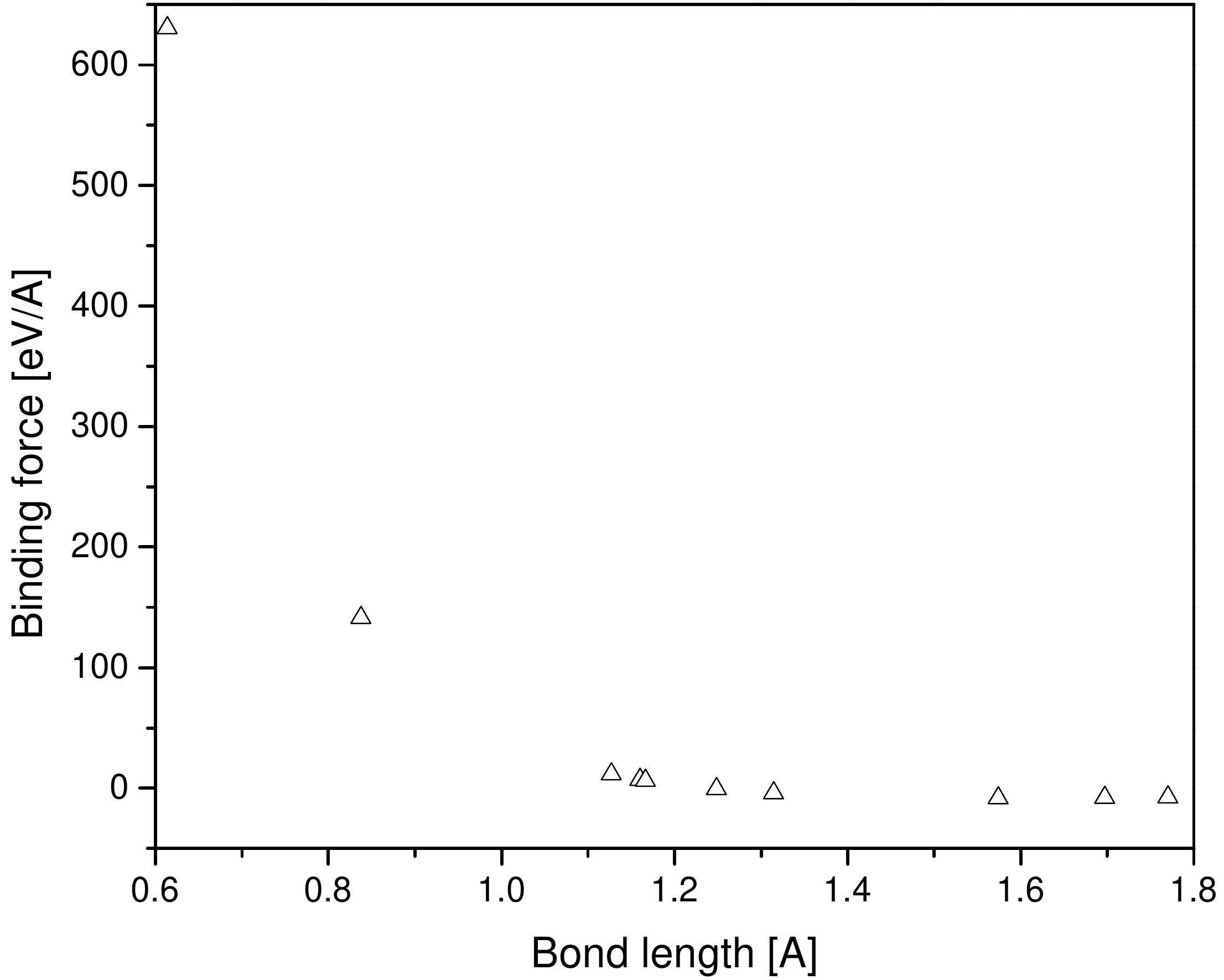}
	\end{subfigure}
	\begin{subfigure}{0.45\textwidth}\caption{} \includegraphics[width=0.99\linewidth]{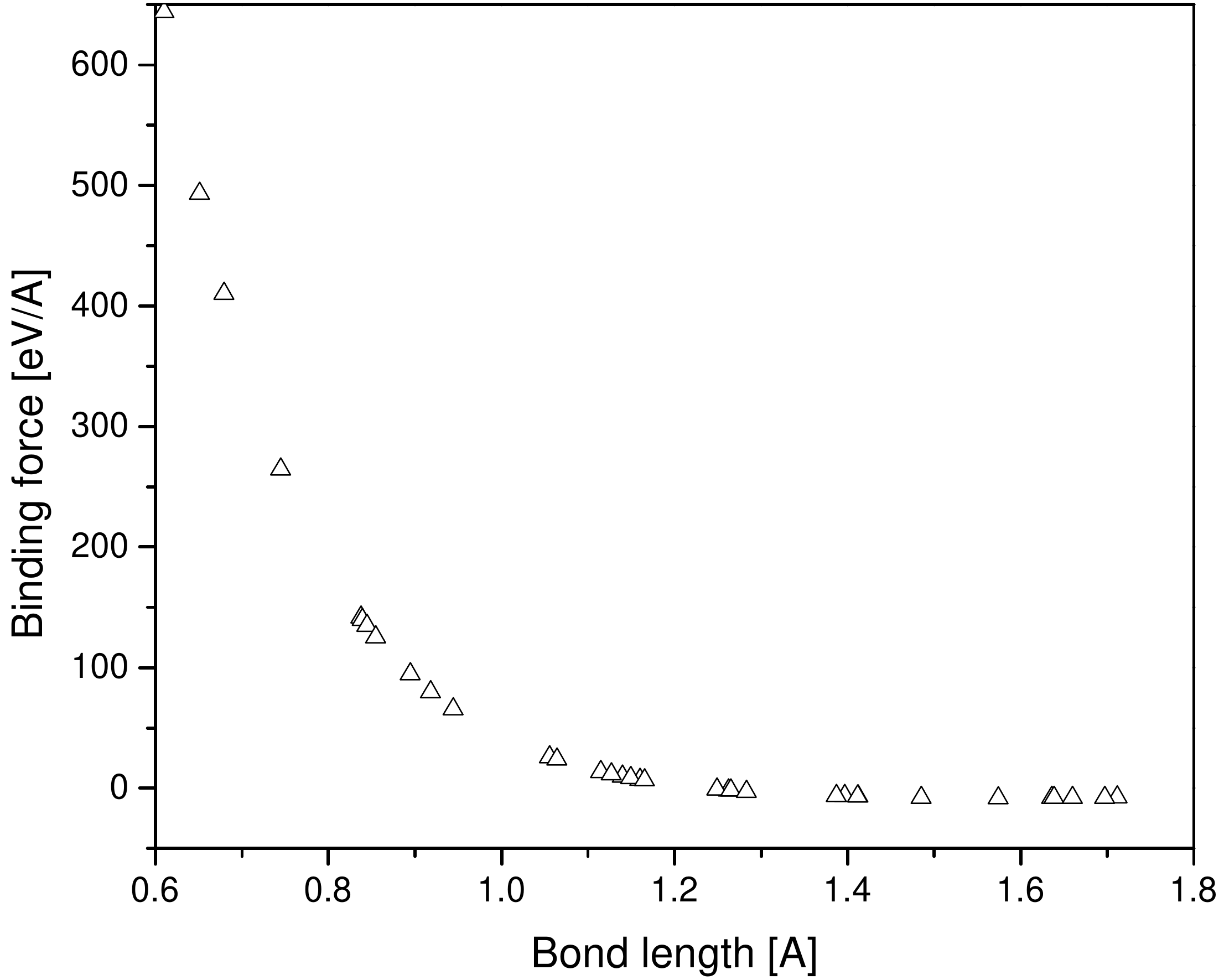}
	\end{subfigure}
	\begin{subfigure}{0.45\textwidth}\caption{} \includegraphics[width=0.99\linewidth]{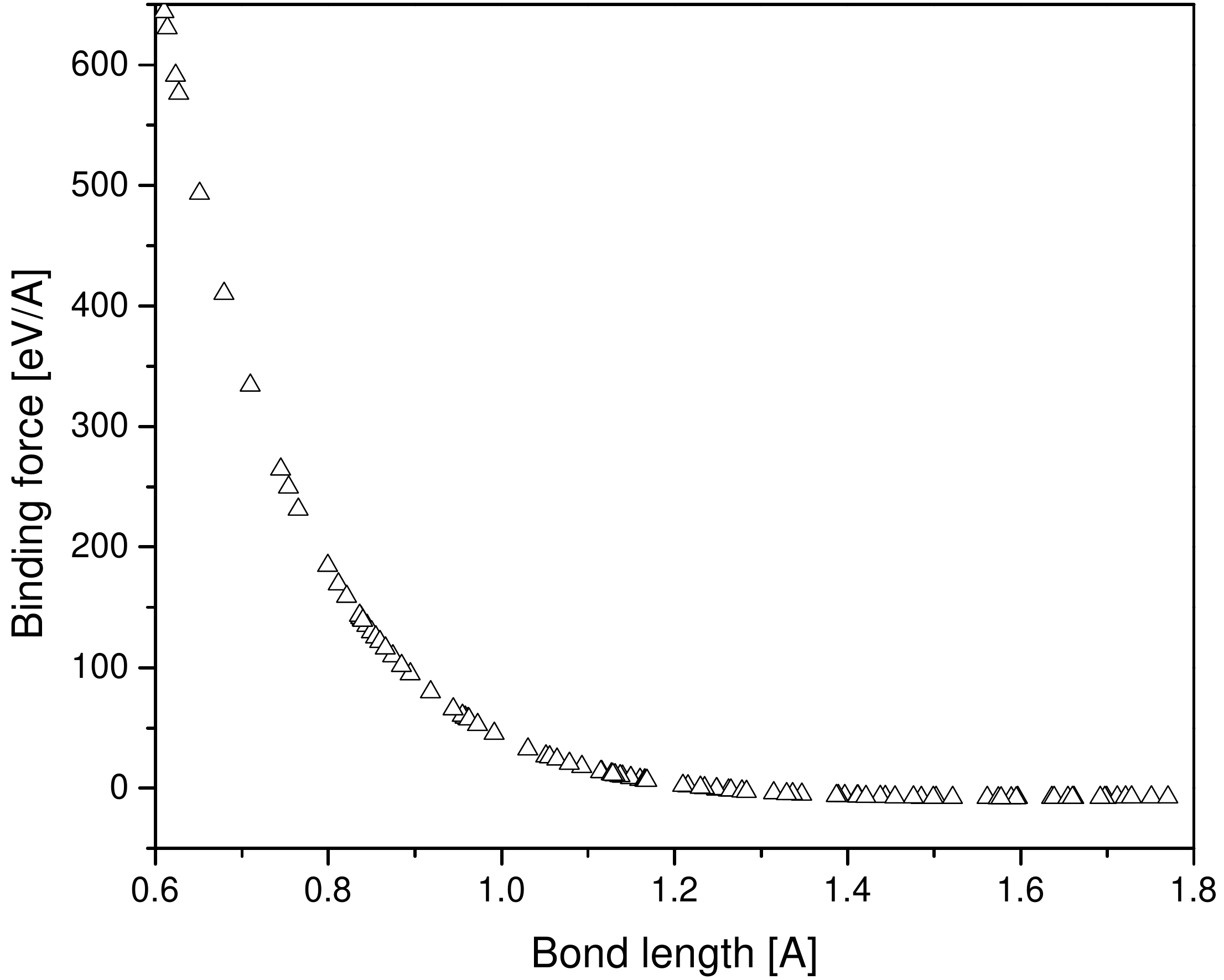}
	\end{subfigure}
    \caption{Sequence of force-field data sets of molecular oxygen O${}_2$ computed using the Vienna {\sl ab initio} Simulation Package (VASP). a) $\#D_1 = 10$ points; b) $\#D_2 = 34$ points; c) $\#D_3 = 100$ points.} \label{atMH3z}
\end{center}
\end{figure}

\begin{example}[Molecular oxygen O${}_2$]\label{kVwY4Z}
{\rm
A simple notional example concerns molecular oxygen O${}_2$, also known as dioxygen or diatomic oxygen, vibrating in a bound configuration without rotation. The ground state of O${}_2$ has a molecular weight of $31.9988$, bond length of $1.21$ {\AA} and a cohesive energy of $-2.58$ eV/atom \cite{CRC:2005}. The oxygen molecule is held together by a strong {O=O} double covalent bond, each oxygen atom sharing two of its outer-shell electrons with the other atom. A sequence of three force-field data sets, containing $\#D_1 = 10$, $\#D_2 = 34$ and $\#D_3 = 100$ points, computed using the Vienna {\sl ab initio} Simulation Package VASP is shown in Fig.~\ref{atMH3z}. The calculations use PAW-PBE as the pseudo-potential and exchange-correlation functional \cite{PhysRevB.50.17953, PhysRevLett.77.3865}. An energy cutoff of 400 eV is employed for the plane wave basis, and reciprocal space is sampled using a $\Gamma$-centered Monkhorst Pack grid \cite{PhysRevB.13.5188}.} \hfill$\square$
\end{example}

The data sets $D_h$ define approximate material trajectory sets of the form
\begin{equation}
     \mathcal{D}_h
    =
    \big\{
        (r_h(\cdot),f_h(\cdot)) \in \mathcal{Z} \, : \,
        (r_h(t),f_h(t)) \in D_h , \; t \in [0,T]
    \big\} .
\end{equation}
Thus, if $D_h$ is a sample of points in phase space, the corresponding material trajectory set $\mathcal{D}_h$ consists of piecewise trajectories that 'visit' the orbits in $D_h$ in turn.

It is clear that, in general, the intersection between the admissible trajectory set $\mathcal{E}$ and the approximating material trajectory sets $\mathcal{D}_h$ may be empty, in which case no classical approximating solution exists. One way to circumvent these difficulty is to relax the notion of 'approximating solutions' generated from the sequence of data sets $D_h$. In particular, we would like such Data-Driven solutions to converge to the exact trajectory as the data sets $D_h$ sample the exact force-field graph $D$ with increasing fidelity.

\subsection{Optimal-control Data-Driven reformulation}\label{7Iw6wu}

In order to derive approximations directly from a sequence of data sets $D_h$, we relax the notion of solution in the spirit of \cite{Kirch:2016, Kirchdoerfer:2017, Kirchdoerfer:2018, conti2018data}to mean a pair $(y_h(\cdot),z_h(\cdot))$ of trajectories in phase space such that $y_h(\cdot) \in \mathcal{D}_h$ is {\sl closest to} $\mathcal{E}$, and, conversely, $z_h(\cdot) \in \mathcal{E}$ is {\sl closest to} $\mathcal{D}_h$, in some metric to be defined. Thus, $y_h(\cdot)$ is the material trajectory that is closest to being admissible and $z_h(\cdot)$ is the admissible trajectory that is closest to being material. Formally,
\begin{equation}\label{vZP0r7}
    \big(y_h(\cdot),z_h(\cdot)\big)
    \in
    {\rm argmin}
    \Big\{
        {\rm dist}\big(y(\cdot),z(\cdot)\big)
        \, : \,
        y(\cdot) \in \mathcal{D}_h , \;
        z(\cdot) \in \mathcal{E}
    \Big\} ,
\end{equation}
where ${\rm dist}$ is some suitable distance between phase-space trajectories, to be defined.

We remark that problem (\ref{vZP0r7}) does not identify, or 'learn', an effective force-field from the data. Instead, it endeavors to minimize discrepancy between admissible solutions, satisfying the equations of motion (\ref{5xvxXr}), and the data, as measured by a suitably defined distance.

Evidently, if the material data set $D_h$ coincides with the entire force-field graph $D$ the solutions of the Data-Driven problem coincide with the exact classical solutions of the initial-value problem (\ref{5xvxXr}), as required. Most importantly, the Data-Driven problem continues to make sense---and returns approximating solutions---in the case of partial point data, thus setting forth an approximation scheme for trajectories. We expect the approximate Data-Driven trajectories to converge to exact classical trajectories as the data samples the force field with increasing fidelity.

We note that problem (\ref{vZP0r7}) has the structure of an {\sl optimal control problem} with differential constraints \cite{Bucur:2005}. To exhibit this structure, suppose, for definiteness, that the distance between trajectories in $\mathcal{Z}$ is of the integral form
\begin{equation}\label{t3FM0D}
    {\rm dist}\big(y(\cdot),z(\cdot)\big)
    =
    \Big(
        \int_0^T {\rm dist}^2(y(t),z(t)) \, dt
    \Big)^{1/2} ,
\end{equation}
for some local distance ${\rm dist}(y,z)$ over $Z$, not renamed. Begin by minimizing out the material trajectories $y(\cdot)$ in (\ref{vZP0r7}) to define a {\sl cost function}
\begin{equation}\label{5u8VbC}
    F_h(z(\cdot))
    =
    \inf_{y(\cdot)\in \mathcal{D}_h}
        \frac{1}{2}
        {\rm dist}^2\big(y(\cdot),z(\cdot)\big)
    =
    \int_0^T
        \frac{1}{2}
        {\rm dist}^2(z(t), D_h )
    \, dt ,
\end{equation}
where
\begin{equation}
    {\rm dist}(z, D_h )
    =
    \inf_{y \in D_h} {\rm dist}(y, z) ,
\end{equation}
is the distance from state $z \in Z$ to the data set $D_h$. Evidently, the cost function $F_h(z(\cdot))$ measures the discrepancy between a trial trajectory $z(\cdot)$ and the force-field data. With these definitions, the abstract Data-Driven problem (\ref{vZP0r7}) reduces to
\begin{subequations}
\begin{align}
    &
    \text{Minimize:} \; F_h(z(\cdot)) \ \text{in } \mathcal{Z},
    \\ & \label{Kk1jWZ}
    \text{subject to: Eqs.~of motion} \; (\ref{5xvxXr}) .
\end{align}
\end{subequations}
As announced, this problem has the structure of an optimal control problem constrained by ordinary-differential equations if, with $z(\cdot) = (r(\cdot),f(\cdot))$, we identify $f(\cdot)$ as the {\sl control} and $r(\cdot)$ as the {\sl state variable} (cf.~\cite{Bucur:2005}, Section~3.5). The problem is then to minimize the cost $F(r(\cdot),f(\cdot))$ with respect to the control $f(\cdot)$, with $r(\cdot)$ taken as a solution of the initial-value problem (\ref{5xvxXr}) for any trial force-field trajectory $f(\cdot)$. General conditions for existence and convergence of solutions for such problems are given in \cite{Bucur:2005}, Section~3.5.

In practice, the constraint (\ref{Kk1jWZ}) can be enforced by means of Lagrange multipliers $w(t)$, subject to initial conditions
\begin{equation}\label{5pCIel}
    w(0) = 0 ,
    \quad
    \dot{w}(0) = 0 ,
\end{equation}
resulting in the Lagrangian
\begin{equation}
\begin{split}
    L\big(y(\cdot),z(\cdot),w(\cdot)\big)
    & =
    \frac{1}{2}
    {\rm dist}^2\big(y(\cdot),z(\cdot)\big)
    \\ & +
    \int_0^T
        \Big(
            m \ddot{r}(t)
            +
            {f}(t)
            -
            f^{\rm ext}(t)
        \Big)
        \, w(t)
    \, dt ,
\end{split}
\end{equation}
to be rendered stationary. Assuming, for definiteness, an Euclidean distance of the form
\begin{equation}
    {\rm dist}^2(y,z)
    =
    \sum_{i=1}^N \Big( | r_i - s_i |^2 + \kappa^2 | f_i - g_i |^2\Big),
\end{equation}
with $y=(s,g)$, $z=(r,f)$,
the Euler-Lagrange equations corresponding to variations in $z(\cdot)$ and $w(\cdot)$ are
\begin{subequations}\label{qhw8ZY}
\begin{align}
    & \label{Q1Z9bs}
    r_h(t) - s_h(t) + m \ddot{w}_h(t) = 0,
    \\ & \label{vq3NEE}
    \kappa^2 (f_h(t) - g_h(t)) + w_h(t) = 0,
    \\ & \label{Yx3eGx}
    m \ddot{r}_h(t) + {f}_h(t) - f_h^{\rm ext}(t) = 0.
\end{align}
\end{subequations}
where we write $y_h(\cdot) = (s_h(\cdot), g_h(\cdot))$ and $z_h(\cdot) = (r_h(\cdot), f_h(\cdot))$ for the resulting Data-Driven solutions. In (\ref{qhw8ZY}), $(s_h(t),g_h(t))$ is to be chosen at all times $t \in [0,T]$ as the point in $D_h$ closest to $(r_h(t),f_h(t))$. We note that, by this choice, problem (\ref{qhw8ZY}) is reduced to two coupled second-order ordinary differential equations in time in the unknowns $r_h(\cdot)$ and $w_h(\cdot)$. This duplicate structure is reminiscent of static Data-Driven problems in which the equilibrium constraint is enforced by means of Lagrange multipliers \cite{Kirch:2016}.

Suppose that $D_h = D$, i.~e., the data supply a full representation of the force field of the system. Then, in (\ref{Q1Z9bs}) and (\ref{vq3NEE}) we have $r_h(t) = s_h(t)$ and $f_h(t) = g_h(t)$, which, together with (\ref{5pCIel}) give $w_h(t) = 0$. In addition, (\ref{Yx3eGx}) reduces to (\ref{0kzJ6C}), as required. In general, for underlying force-field graphs of sufficient regularity we expect the Lagrange multipliers $w_h(\cdot)$ to tend to zero and the Data-Driven trajectories to converge to exact classical trajectories when the density of sampling increases and the data sets $D_h$ approximate $D$ with increasing fidelity.

\subsection{Game-theoretical Data-Driven reformulation}\label{4iH7kb}

An alternative Data-Driven paradigm that does learn an effective force-field from the data consists of recasting the Data-Driven problem (\ref{vZP0r7}) as a game-theoretical problem. To this end, we regard the functional $-F_h(r(\cdot),f(\cdot))$, eq.~\ref{5u8VbC} with $z(\cdot) = (r(\cdot),f(\cdot))$, as the payoff for the {\sl force player}. Evidently, $-F_h(r(\cdot),f(\cdot))$ which measures the agreement between the trajectory $z(\cdot) = (r(\cdot),f(\cdot))$ and the data set. In this reinterpretation, the {\sl objective} of the force player is to determine a {\sl strategy} $f(\cdot)$ that maximizes its payoff, or, equivalently, minimizes $F_h(r(\cdot), \cdot)$, for given $r(\cdot)$. The force player competes against a second {\sl position player}, whose {\sl objective} is to determine a {\sl strategy} $r(\cdot)$ that satisfies the initial-value problem (\ref{5xvxXr}) for given $f(\cdot)$. The objective of the position player can be expressed variationally by means of the action functional
\begin{equation}
    G(r(\cdot),f(\cdot))
    =
    \int_0^T
        \sum_{i=1}^N
        \Big(
            \frac{m_i}{2} |\dot{r}_i(t)|^2
            -
            \big( f_i(t) - f_i^{\rm ext}(t) \big) \cdot r_i(t)
        \Big)
    \, dt ,
\end{equation}
to be minimized with respect to $r(\cdot)$ at fixed $f(\cdot)$, with initial conditions (\ref{vKAl5W}) replaced by the boundary conditions
\begin{equation}\label{0NXBqI}
    r(0) = r_0 , \quad r(T) = r_T ,
\end{equation}
with $r_0$ and $r_T$ given. The minimizing property of the action functional, which normally only attains stationarity in general Lagrangian mechanics, is remarkable and owes to the independence of the force field. The Data-Driven problem then becomes
\begin{subequations}\label{1qsNJ6}
\begin{align}
    &   \label{q9UNx4}
    f_h(\cdot)
    \in
    \mathop{\rm argmin} F_h(r_h(\cdot),\cdot) ,
    \\ &
    r_h(\cdot)
    \in
    \mathop{\rm argmin} G(\cdot, f_h(\cdot)) ,
\end{align}
\end{subequations}
with defines a non-cooperative game between the force and position players \cite{Roubicek:2020}.

The difference between the game (\ref{1qsNJ6}) and the optimal control problem (\ref{vZP0r7}) is subtle but significant. Thus, in the optimal control problem the trial position histories $r(\cdot)$ are tied to the trial control histories $f(\cdot)$ through the initial-value problem (\ref{5xvxXr}). By virtue of this constraint, the cost functional $F_h(r(\cdot), f(\cdot))$ has a double dependence on the trial control histories $f(\cdot)$, once through its direct dependence and twice implicitly through $r(\cdot)$. By contrast, in the game problem (\ref{1qsNJ6}) the same functional $F_h(r(\cdot), f(\cdot))$ is minimized with respect to $f(\cdot)$ at fixed $r(\cdot)$. Therefore, the Data-Driven approximations $(r_h(\cdot), f_h(\cdot))$ generated by the two procedures are different in general.

Specifically, suppose that we fix a trial trajectory $r(\cdot)$ in configuration space. From (\ref{5u8VbC}) and (\ref{q9UNx4}) if follows that the optimal force strategy $f(\cdot)$ of the force player is given at every time by the {\sl effective force field}
\begin{equation}
    f(r(t); D_h) = f(t) \, : \,
    (s(t),f(t)) \in D_h , \;
    {\rm dist}(s(t),r(t)) \to \min! ,
\end{equation}
which assigns to every system position $r(t)$ the force $f(t)$ such that $(s(t),f(t)) \in D_h$ and $s(t)$ is nearest to $r(t)$. The optimal position strategy $r_h(\cdot)$ then follows by solving the initial-value problem
\begin{subequations}\label{8mCJjz}
\begin{align}
    &
    m_i \ddot{r}_i(t) + f_i(r(t); D_h) = f_i^{\rm ext}(t) ,
    \\ &
    r(0) = r_0,
    \quad
    \dot{r}(0) = v_0 ,
\end{align}
\end{subequations}
formally identical to (\ref{5xvxXr}) with the exact, but unknown, force-field $f(r)$ replaced by the 'learned' Data-Driven force field $f(r; D_h)$.

It bears emphasis that the 'learned' Data-Driven force field $f(r; D_h)$ is not the product of modeling, but rather a result of analysis. Piecewise constant approximations of the force field have been analyzed by Gonzalez {\sl et al.} \cite{Gonzalez:2010}, who termed the approximation scheme {\sl force stepping} and analyzed its convergence properties. As in the optimal control formulation, we expect the approximate trajectories $(r_h(\cdot), f_h(\cdot))$ obtained from (\ref{8mCJjz}) to converge to the exact solution as the data sets $D_h$ approximate the graph $D$ with increasing fidelity. General conditions for existence and convergence of solutions of general game-theoretical problems are given in \cite{Roubicek:2020}.

\subsection{Convergence with respect to the data}\label{qQV1fx}

The convergence properties of the Data-Driven solutions $(r_h(\cdot), f_h(\cdot))$ to the exact trajectory $(r(\cdot), f(\cdot))$ as the material data set $D_h$ samples the force-field graph $D$ with increasing fidelity can easily be verified for simple scenarios. The analysis is greatly simplified by assuming long-term existence of trajectories $(r(\cdot), f(\cdot))$ and then focusing on the convergence of approximations $(r_h(\cdot), f_h(\cdot))$ thereof, in keeping with the main focus of this work.

We recall that a function $f:\mathbb{R}^m \to \mathbb{R}^n$ is Lipschitz continuous if there is a constant $L>0$ such that
\begin{equation}
    \| f(r') - f(r'') \| \leq L \, \| r' - r ''\| .
\end{equation}
for all $r'$, $r'' \in \mathbb{R}^m$. The space of Lipschitz-continuous functions of time over $[0,T]$ with values in $\mathbb{R}^n$ is denoted $W^{1,\infty}((0,T);\mathbb{R}^n)$; it is a Banach space with norm
\begin{equation}
    \| r \|_{1,\infty}
    =
    \max
    \big\{
        \mathop{\rm ess\sup}_{t\in(0,T)}{|r(t)|},
        \; T
        \mathop{\rm ess\sup}_{t\in(0,T)}{|\dot{r}(t)|}
    \big\} ,
\end{equation}
where ${\rm ess\sup}$ denotes the essential supremum (cf., e.~g., \cite{Evans:1998}).

\begin{prop}[Convergence of optimal-control Data-Driven problem]\label{bwh2P1}
Suppose that $f(r)$ is Lipschitz continuous and the exact initial-value problem (\ref{5xvxXr}) has solutions $r(\cdot)$ in $W^{1,\infty}((0,T); \mathbb{R}^{3N})$. Suppose that the data sets $D_h$ are generated by sampling the exact force field $f(r)$ at points $s \in \mathbb{R}^{3N}$ and that there is $\epsilon_h \downarrow 0$ such that for every $r \in \mathbb{R}^{3N}$ there is $(s,g) \in D_h$ with $| r-s | \leq \epsilon_h$. Then, the optimal-control Data-Driven solutions $r_h(\cdot)$ converge to the exact solution $r(\cdot)$ strongly in $W^{1,\infty}((0,T),\mathbb{R}^{3N})$.
\end{prop}

The assumption of Lipschitz continuity places restrictions on the force field. For instance, for a dimer such as molecular oxygen O${}_2$, Example~\ref{kVwY4Z}, Lipschitz continuity and inversion symmetry require $f(r)$ to be continuous through the origin $r=0$ with $f(0) = 0$, i.~e., the interaction force must vanish when the two particles merge. We also recall that, if $f(r)$ is differentiable, then the Lipschitz constant $L$ is given by the maximum value of $|Df(r)|$, where $Df(r)$ is the matrix of partial derivatives of $f(r)$, or Hessian. Convergence in  $W^{1,\infty}$ means, in particular, that trajectories $r(\cdot)$ in configuration space are Lipschitz continuous and the velocities $\dot{r}(\cdot)$ exist almost everywhere in $[0,T]$ and are essentially bounded.

The convergence of the game-theoretical Data-Driven solutions can be verified likewise.

\begin{prop}[Convergence of game-theory Data-Driven problem]\label{dN4PE6}
Under the assumptions of Prop.~\ref{bwh2P1}, the game-theoretical Data-Driven solutions $r_h(\cdot)$ converge to the exact solution $r(\cdot)$ strongly in $W^{1,\infty}((0,T),\mathbb{R}^{3N})$.
\end{prop}

The proofs of the preceding propositions are straightforward and are presented in Appendix A for completeness.

\subsection{The distance between $E(3)$-orbits of point sets}\label{ZJ1cqZ}

\begin{figure}[ht]
\begin{center}
	\includegraphics[width=0.575\linewidth]{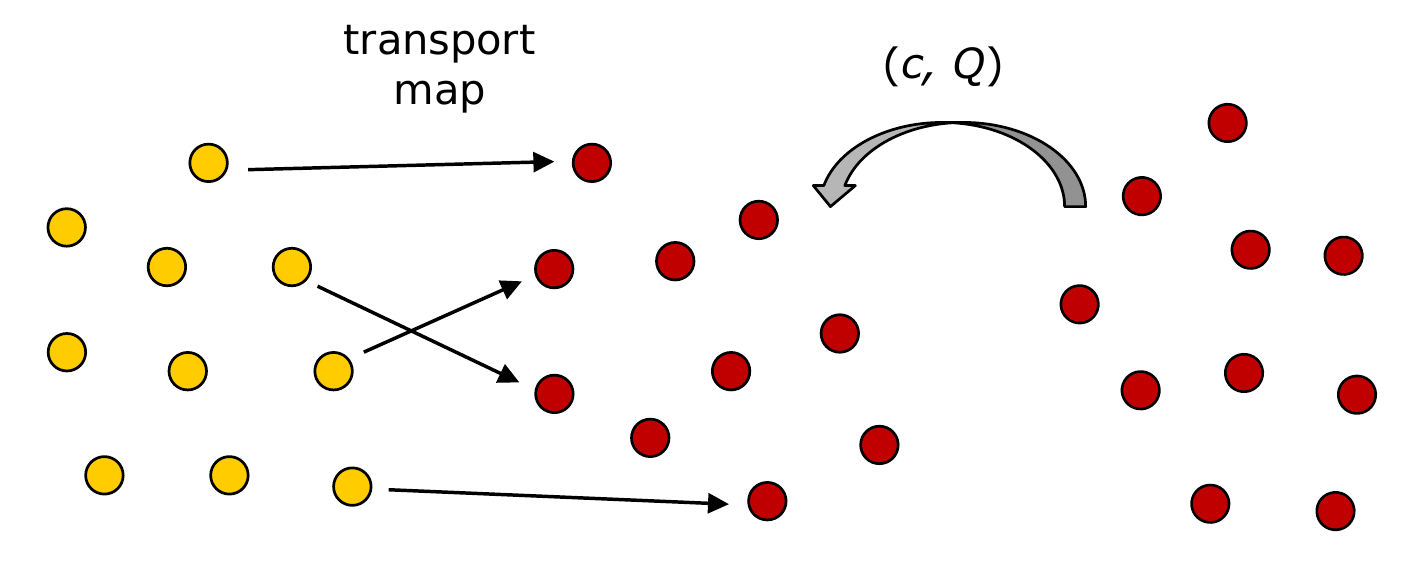}
    \caption{Schematic representation of the Wasserstein distance between the $E(n)$-orbits of two clusters of points (yellow and red).} \label{0OwXeo}
\end{center}
\end{figure}

Recall that the force-field graph $D$ and material data sets $D_h$ thereof are invariant under the action of the Euclidean group $E(3)$, therefore contain entire $E(3)$-orbits, and are additionally invariant under permutations of the atoms. However, in practice actual material data sets are likely to consist of representatives of the $E(3)$-orbits pegged to a particular numbering of the atoms. Details of the calculation of distances between $E(3)$-orbits of point sets from representatives thereof are presented in this section.

Distances between point sets are naturally measured by the discrete Wasserstein distance \cite{Evans:1997}, i.~e., the cost of transportation of one point set to another. Thus, let $r = \{r_i\}_{i=1}^{N}$ and $s = \{s_i\}_{j=1}^N$ be two clusters of points in configuration space $\mathbb{R}^{3N}$ (e.~g., red and yellow in Fig.~\ref{0OwXeo}) and let $f = \{f_i\}_{i=1}^{N}$ and $g = \{g_i\}_{j=1}^N$ be force fields attached to them, respectively. We write $y=(s,g)$ and $z=(r,f)$ for the corresponding points in phase space. For simplicity, we consider clusters with the same number of points, though the optimal transport framework can be extended to clusters of different size \cite{Evans:1997}. The discrete $2$-Wasserstein distance between the $E(3)$-orbits of the point sets $\cup_{i=1}^N\{y_i\}$ and $\cup_{i=1}^N\{z_i\}$ in $\mathbb{R}^3\times \mathbb{R}^3$ is
\begin{equation}\label{sja6Ga}
    {\rm dist}^2(y,z)
    =
    \min_{\sigma \in S_N, \, (c,Q)\in E(3)}
    \sum_{i=1}^N \Big( | r_i - Q(s_{\sigma(i)}-c)|^2 + \kappa^2 | f_i - {Q} g_{\sigma(i)} |^ 2\Big),
\end{equation}
where the minimum is sought over the group $S_N$ of all permutations $\sigma$ of the index set $\{1,\dots,N\}$, translations $c \in \mathbb{R}^3$ and orthogonal transformations $Q \in O(3)$, Fig.~\ref{0OwXeo}. Fixing $\sigma \in S_N$ and taking variations of (\ref{sja6Ga}) with respect to $(c,Q)$ gives the Euler-Lagrange equations
\begin{subequations}\label{7WajHF}
\begin{align}
    & \label{Ege4ov}
    \sum_{i=1}^N \big( {Q}^{-1} r_i - (s_{\sigma(i)}-c) \big) = 0 ,
    \\ & \label{r4EPTr}
    \sum_{i=1}^N
    {Q}^{-1}
    \big( r_i \otimes (s_{\sigma(i)}-c)  + \kappa^2 f_i \otimes g_{\sigma(i)} \big)
    = U = U^T .
\end{align}
\end{subequations}
From (\ref{Ege4ov}), we obtain
\begin{equation}\label{uzV7pZ}
    c = \frac{1}{N} \sum_{i=1}^N \big( {Q}^{-1} r_i - s_{\sigma(i)} \big) ,
\end{equation}
which, inserted into (\ref{r4EPTr}), gives
\begin{equation}
    \sum_{i=1}^N
    {Q}^{-1}
    \Big(
        r_i \otimes s_{\sigma(i)}
        +
        r_i \otimes
        \big(
        \frac{1}{N} \sum_{j=1}^N s_{\sigma(j)} \big)
        +
        \kappa^2
        f_i \otimes g_{\sigma(i)}
    \Big)
    =
    U
    =
    U^T ,
\end{equation}
upon a redefinition of $U$. Evidently, the orthogonal transformation $Q$ results from the polar decomposition of a $3\times 3$ matrix that is explicit in the data. Once, $Q$ is determined, the translation $c$ follows from (\ref{uzV7pZ}), and the corresponding trial distance as
\begin{equation}
    {\rm dist}_\sigma^2(y,z)
    =
    \sum_{i=1}^N \Big( | r_i - Q (s_{\sigma(i)}-c)|^2 + \kappa^2 | f_i - {Q} g_{\sigma(i)} |^ 2\Big) .
\end{equation}
The sought distance (\ref{sja6Ga}) then follows by minimizing over the trial distances from all permutations of the indices.

Evidently, the exact evaluation of the distance (\ref{sja6Ga}) is of combinatorial complexity due to the need to examine all permutations of the atoms and it is suitable only for small clusters. An alternative form of the distance is the Kantorovich reformulation \cite{Evans:1997}
\begin{equation}\label{CH6wxk}
\begin{split}
    &
    {\rm dist}^2(r,s)
    =
    \inf_{\begin{array}{cc}\mu \in \mathbb{R}^{{N}\times N} \\ (c,Q)\in E(3) \end{array}}
    \Big\{
        \sum_{i=1}^{N}
        \sum_{j=1}^N
        \mu_{ij}
        \Big(
            | r_i - Q(s_j-c) |^2
            + \\ &
            \kappa^2 | f_i - Q g_j |^2
        \Big) \, : \,
        \ \mu_{ij} \geq 0,
        \ \sum_{j=1}^N \mu_{ij} = 1
    \Big\} ,
\end{split}
\end{equation}
which reduces the computation to the solution linear programming problem for the weights $\mu_{ij}$.

Explicit approximations for the weights can further be obtained by recourse to interior point methods \cite{ShuCherng:2001}. For instance, a maximum-entropy (max-ent) regularization \cite{ArroyoOrtiz2006} consists of replacing (\ref{CH6wxk}) by
\begin{equation}
\begin{split}
    &
    {\rm dist}^2(r,s)
    =
    \inf_{\begin{array}{cc}\mu \in \mathbb{R}^{{N}\times N} \\ (c,Q)\in E(3) \end{array}}
    \Big\{
        \sum_{i=1}^{N}
        \sum_{j=1}^N
        \mu_{ij}
        \Big(
            | r_i - Q(s_j-c) |^2
            + \\ &
            \kappa^2 | f_i - Q g_j |^2
            +
            \frac{1}{\beta} \log(\mu_{ij})
        \Big)
        \, : \,
        \ \sum_{j=1}^N \mu_{ij} = 1
    \Big\} ,
\end{split}
\end{equation}
which can be readily solved for the approximate weights $\mu^\beta_{ij}$. A further minimization with respect to $(c,Q)$, exploiting the optimality of $\mu^\beta_{ij}$, results in Euler-Lagrange equations identical to (\ref{7WajHF}). The $2$-Wasserstein distance ${\rm dist}(y,z)$ then follows from the regularized, or {\sl thermalized}, distance ${\rm dist}_\beta(y,z)$ in the limit of $\beta \to +\infty$.

We note that the game-theoretical Data-Driven problem requires distances in configuration space only. Such distances follow from the preceding expressions as a special case by formally setting $\kappa = 0$.

\subsection{Time discretization}\label{He0WpI}

Suppose now that we wish to approximate trajectories at discrete points $t_0=0, t_1,\dots,t_n,t_{n+1}=t_n+\tau,\dots,T$, where, for simplicity, we consider a constant time step $\tau$ with integer $T/\tau$. The discrete trajectories are then sequences $(r_\cdot,f_\cdot) \equiv (r_n,f_n)_{n=0}^{T/\tau}$ of points in phase space. We denote by $\mathcal{Z}_\tau$ the space of such discrete phase-space trajectories.

{\sl NB: As in the time-continuous case, Section~\ref{N3q49y}, in order to carefully differentiate between points and discrete trajectories, henceforth we shall denote by $r_\cdot$, $f_\cdot$ and $z_\cdot=(r_\cdot,f_\cdot)$ trajectories in $\mathcal{Z}_\tau$. In particular, $r_n$, $f_n$ and $z_n=(r_n,f_n)$ denote the values of trajectories $r_\cdot$, $f_\cdot$ and $z_\cdot=(r_\cdot,f_\cdot)$  at time $t_n = n \tau \in [0,T]$, respectively.}

In the time-discrete setting, we may approximate the equations of motion by means of a general difference formula. For definiteness, we consider three-point formuale of the form
\begin{equation}\label{Z3qX77}
\begin{split}
    &
    m_i \frac{r_i^{n+1} + r_i^{n-1}- 2 r_i^n}{\tau^2}
    +
    \alpha_{-1}\big(f_i^{n-1} - f_i^{\rm ext}(t_{n-1}) \big)
    + \\ &
    \alpha_0\big(f_i^n - f_i^{\rm ext}(t_n) \big)
    +
    \alpha_1\big(f_i^{n+1} - f_i^{\rm ext}(t_{n+1}) \big)
    =
    0 ,
\end{split}
\end{equation}
with
\begin{equation}
    \alpha_{-1} + \alpha_0 + \alpha_1 = 1 ,
\end{equation}
and denote by $\mathcal{E}_\tau$ the set of discrete trajectories satisfying the discrete equations of motion (\ref{Z3qX77}) and initial conditions, namely,
\begin{equation}
    \mathcal{E}_\tau
    =
    \big\{
        z_. = (r_.,f_.) \in \mathcal{Z}_\tau \, : \, (\ref{Z3qX77}), \; (r_0,r_1) \; \text{given}
    \big\} .
\end{equation}
Likewise, we identify the set of discrete material trajectories as
\begin{equation}
    \mathcal{D}_\tau
    =
    \big\{
        y_\cdot = (r_\cdot,f_\cdot) \in \mathcal{Z}_\tau \, : \, (r_n,f_n) \in D, \; n=1,\dots,T/\tau
    \big\} ,
\end{equation}
and the set of discrete material-data trajectories as
\begin{equation}
    \mathcal{D}_{h,\tau}
    =
    \big\{
        y_\cdot = (r_\cdot,f_\cdot) \in \mathcal{Z}_\tau \, : \, (r_n,f_n) \in D_h, \; n=1,\dots,T/\tau
    \big\} .
\end{equation}

The time-discrete versions of the optimal-control and game-theoretical Data-Driven formulations defined in the foregoing follow now {\sl mutatis mutandi} through the introduction of a sui metric in the space $\mathcal{Z}_\tau$ of discrete phase-space trajectories. For instance, we may replace (\ref{t3FM0D}) by
\begin{equation}
    {\rm dist}_\tau\big(y_\cdot,z_\cdot\big)
    =
    \Big(
        \sum_{n=0}^{T/\tau} {\rm dist}^2(y_n,z_n) \, \tau
    \Big)^{1/2} .
\end{equation}
Then, the time-discrete version of the optimal-control Data-Driven problem (\ref{vZP0r7}) is
\begin{equation}
    \big(y^{h,\tau}_\cdot,z^{h,\tau}_\cdot\big)
    \in
    {\rm argmin}
    \Big\{
        {\rm dist}_\tau\big(y_\cdot,z_\cdot\big)
        \, : \,
        y_\cdot \in \mathcal{D}_{h,\tau} , \;
        z_\cdot \in \mathcal{E}_\tau
    \Big\} .
\end{equation}
As in the time-continuous case, the constraint set forth by the discrete equations of motion (\ref{Z3qX77}) can be enforced by means of a discrete Lagrange multiplier $w_\cdot$ satisfying homogeneous initial conditions. Stationarity then results in two coupled second-order recurrence relations for $r_\cdot$ and $w_\cdot$.

Likewise, a time-discrete version of the game-theoretical Data-Driven problem (\ref{1qsNJ6}) can be set forth by introducing the
discrete cost function, cf.~eq.~(\ref{5u8VbC}),
\begin{equation}
    F_{h,\tau}(z(\cdot))
    =
    \inf_{y_\cdot \in \mathcal{D}_{h,\tau}}
        \frac{1}{2}
        {\rm dist}_\tau^2\big(y_\cdot,z_\cdot\big)
    =
    \sum_{n=0}^{T/\tau}
        \frac{1}{2}
        {\rm dist}^2(z_n, D_h )
    \, \tau .
\end{equation}
The objective of the force player is to determine a strategy $f_\cdot$ that maximizes its payoff, or, equivalently, minimizes $F_{h,\tau}(r_\cdot, \cdot)$, for given $r_\cdot$. The force player competes against a second {\sl position player}, whose objective is to determine a strategy $r_\cdot$ that satisfies the discrete equations of motion (\ref{Z3qX77}) for fixed $f_\cdot$. The objective of the position player can be expressed variationally by means of the discrete action functional
\begin{equation}
    G_\tau(r_\cdot,f_\cdot)
    =
    \sum_{n=0}^{T/\tau}
        \sum_{i=1}^N
        \Big(
            \frac{m_i}{2} |\dot{r}_i^n|^2
            -
            \big( f_i^n - f_i^{\rm ext}(t_n) \big) \cdot r_i^n
        \Big)
    \, \tau
\end{equation}
to be minimized with respect to $r_\cdot$ at fixed $f_\cdot$ with given $r_0$ and $r_{T/\tau}$.

The treatment of the time-discrete problems is otherwise identical to that of the time-continuous problems and details are therefore omitted in the interest of brevity.

Assuming that the time discretization scheme is strongly convergent, i.~e., time-discrete trajectories converge to the exact time-continuous trajectories strongly in $W^{1,\infty}((0,T),\mathbb{R}^{3N})$ as $\tau \downarrow 0$,
the joint convergence of the time-discrete Data-Driven trajectories with respect to data and time discretization follows directly from Props.~\ref{bwh2P1} and \ref{dN4PE6} by passing to diagonal sequences.

\subsection{Data-driven Verlet algorithm}\label{D63kfb}

\begin{algorithm}[H]
\caption{Data-driven Verlet algorithm}
\label{p4ZuLW}
\begin{algorithmic}
\REQUIRE
Number of atoms $N$; $r_{n-1}$, $r_n$ and $f_n$; force-field data set $D_h$; applied loads $f_n^{\rm ext}$. Then:
\STATE i) Compute $r_{n+1}$ from (\ref{t7BGhs}).
\STATE iii) Find $(s_{n+1},g_{n+1})$ in $D_h$ such that $s_{n+1}$ is closest to $r_{n+1}$.
\STATE iii) Set $f_{n+1} = g_{n+1}$.
\STATE Return $r_n$, $r_{n+1}$ and $f_{n+1}$.
\end{algorithmic}
\end{algorithm}

An explicit time-discretization scheme commonly used in practice is {\sl Verlet's algorithm}
\begin{equation}\label{t7BGhs}
    m_i \frac{r_i^{n+1} + r_i^{n-1}- 2 r_i^n}{\tau^2}
    +
    f_i^n
    =
    f_i^{\rm ext}(t_n) ,
\end{equation}
which is a special case of (\ref{Z3qX77}) with $\alpha_{-1} = \alpha_1 = 0$. It is readily verified that, in this case, the optimal control and game-theoretical Data-Driven problems coincide.

The resulting time-stepping scheme is summarized in Algorithm~\ref{p4ZuLW}. It bears emphasis that the Data-Driven Verlet algorithm retains the explicit character of the classical Verlet algorithm. The determination of $(r_{n+1},f_{n+1})$ requires a search over the force-field data set $D_h$, which is the main computational bottleneck of the algorithm. Fast algorithms for searching large data sets may be found in \cite{Eggersmann:2021} and references therein.

\subsection{Short-ranged force fields}

As already remarked, the computation of distances between $E(3)$-orbits of point sets is computationally intensive for large systems. Conveniently, many force fields of practical interest are {\sl short-ranged}, which allows distance calculations to be restricted to small clusters defined according to the range of interaction.

Suppose, for simplicity, that the system under consideration is conservative and the attendant global force field $f(r)$ derives from a global potential $V(r)$. Suppose, in addition, that $V(r)$ is short-ranged. In order to exploit this property, we define the local interatomic potentials
\begin{equation}\label{k7Sr3x}
    V_i(r)
    =
    \int
        f_i(r)
    \cdot dr_i
    =
    \int
        \frac{\partial V}{\partial r_i}(r)
    \cdot dr_i ,
\end{equation}
in terms of indefinite integrals defined modulo additive constants. Thus, $V_i(r)$ is the primitive function of $f_i(r)$ with respect to $r_i$ with the coordinates of all other atoms held constant. We note that, by the definition of the local interatomic potentials, we have
\begin{equation}\label{aSUh5E}
    f_i(r)
    =
    \frac{\partial V_i}{\partial r_i}(r) ,
\end{equation}
i.~e., the force at atom $i$ follows as the derivative of the local interatomic potential $V_i$ with respect to $r_i$, with the coordinates of all other atoms held constant.

It bears emphasis that definition (\ref{k7Sr3x}) applies to arbitrary interatomic potentials without loss of generality. In addition, from (\ref{aSUh5E}) it follows that the collection of local interatomic potentials jointly supply the exact global force field of the system and, therefore, it affords an equivalent representation of said global force field.

For short-ranged interatomic potentials, we may expect the number of atoms involved in the definition of each local interatomic potentials to be much smaller than $N$. Two atoms $i$ and $j$ are {\sl non-interacting} if
\begin{equation}\label{8MJLzR}
    \frac{\partial f_i}{\partial r_j}(r)
    =
    \frac{\partial^2V}{\partial r_i \partial r_j}(r)
    =
    0 ,
\end{equation}
and {\sl interacting} otherwise. We denote by $N_i \subset \{1,\dots,N\}$ the subset of atoms that interact with $r_i$, not including $r_i$ itself, and by $\#N_i$ the number of atoms in $N_i$. It then follows from (\ref{aSUh5E}) that $f_i(r)$ depends only on $r_i$ and its cohort $N_i$ of interacting atoms and that $V_i(r)$ can be chosen to have the same dependence up to inconsequential additive terms.

\begin{example}[Monatomic chain]
{\rm
Consider a monatomic chain consisting of atoms interacting through the harmonic potential
\begin{equation}\label{dk7ZFf}
    V(r) =
    \sum_i \frac{m\omega^2}{2} (r_{i+1}-r_i)^2 .
\end{equation}
The local potentials satisfy the relations
\begin{equation}
    \frac{\partial V_i}{\partial r_i}(r)
    =
    \frac{\partial V}{\partial r_i}(r)
    =
    m\omega^2 (r_i-r_{i-1})
    +
    m\omega^2 (r_i-r_{i+1}) .
\end{equation}
Integrating, we obtain
\begin{equation}
    V_i(r)
    =
    \frac{m\omega^2}{2} (r_i-r_{i-1})^2
    +
    \frac{m\omega^2}{2} (r_{i+1}-r_i)^2  ,
\end{equation}
up to inconsequential additive constants. We note that, whereas $V(r)$ depends on the coordinates of all atoms, $V_i(r)$ depends only on the coordinate $r_i$ of the central atom and the coordinates of its nearest-neighbors $N_i = \{i-1, i+1\}$.} \hfill$\square$
\end{example}

We proceed to verify that, if the interatomic potential $V(r)$ is invariant under the action of the Euclidean group $E(3)$, then so are the local interatomic potentials $V_i(r)$. To this end, we note from definition (\ref{k7Sr3x}) that, modulo inconsequential additive constants,
\begin{equation}
    V_i(r)
    =
    \int_0^1 \frac{\partial V}{\partial r_i}(\gamma(s)) \cdot \gamma'(s) \, ds
\end{equation}
for any curve $\gamma(s)$ joining a reference configuration $r_0$ to $r$, i.~e., such that $r_0 = \gamma(0)$ and $r = \gamma(1)$. Since, for any $c \in \mathbb{R}^3$ and $Q \in O(3)$, the curve $Q \, \gamma(s) + c$ joints the reference configuration $Q \,  r_0 + c$ to $Q \,  r + c$, it follows that, modulo inconsequential additive constants,
\begin{equation}
    V_i(Q \,  r + c)
    =
    \int_0^1
        \frac{\partial V}{\partial r_i}(Q \,  \gamma(s)+c)
        \cdot
        (Q \,  \gamma'(s))
    \, ds .
\end{equation}
Suppose now that $V(r)$ is invariant under rigid-body mappings. Then,
\begin{equation}
    V_i(Q \,  r + c)
    =
    \int_0^1 \frac{\partial V}{\partial r_i}(\gamma(s)) \cdot \gamma'(s) \, ds
    =
    V_i(r) ,
\end{equation}
modulo inconsequential additive constants, which establishes the invariance of $V_i(r)$.

Finally, we note that the global initial-value problem (\ref{5xvxXr}) may be regarded as a {\sl game} involving $N$ players, one per atom in the system, with strategies $r_i(t)$ whose payoff is to satisfy their corresponding equations of motion (\ref{0kzJ6C}) and initial conditions. In variational form, each player seeks to render stationary its action functional
\begin{equation}
    A_i(r_i(\cdot))
    =
    \int_0^T
        \Big(
            \frac{m_i}{2} | \dot{r}_i(t) |^2
            -
            V_i(r(t)) + f_i^{\rm ext}(t) \cdot r_i(t)
        \Big)
    \, dt ,
\end{equation}
with appropriate boundary conditions. The objective of the $i$th player in the game is then to determine a strategy $r_i(\cdot)$ that renders $A_i(r_i(\cdot))$ stationary. We note that, by (\ref{aSUh5E}), the Euler-Lagrange equations of the game so defined are identical to (\ref{0kzJ6C}), which shows the equivalence between the two paradigms.

Evidently, the Data-Driven reformulation of the global molecular-dynamics problem set forth in the foregoing can be applied, {\sl mutatis mutandi}, to each of the local problems. In order to formulate the local Data-Driven problems, it suffices to supply {\sl local} force-field data sets $D_{h,i}$ collecting corresponding values of $(r_i, r_{|N_i}, f_i)$, i.~e., local position $r_i$ of atom $i$, positions $r_{|N_i}$ of the atoms in the local neighborhood $N_i$ of interaction and corresponding interaction force $f_i$ on atom $i$. The dimensionality of the local material data sets $D_{h,i}$ is thus much reduced with respect to that of the global data set $D_h$ and, in particular, is independent of the global size $N$ of the system.

\section{Numerical examples}

Finally, we present examples of application aimed at demonstrating the feasibility, range and scope of the Data-Driven molecular dynamics paradigm.

\subsection{Diatomic oxygen ${\rm O}_2$}\label{49azWB}

\begin{figure}[ht]
\begin{center}
	\begin{subfigure}{0.45\textwidth}\caption{} \includegraphics[width=0.99\linewidth]{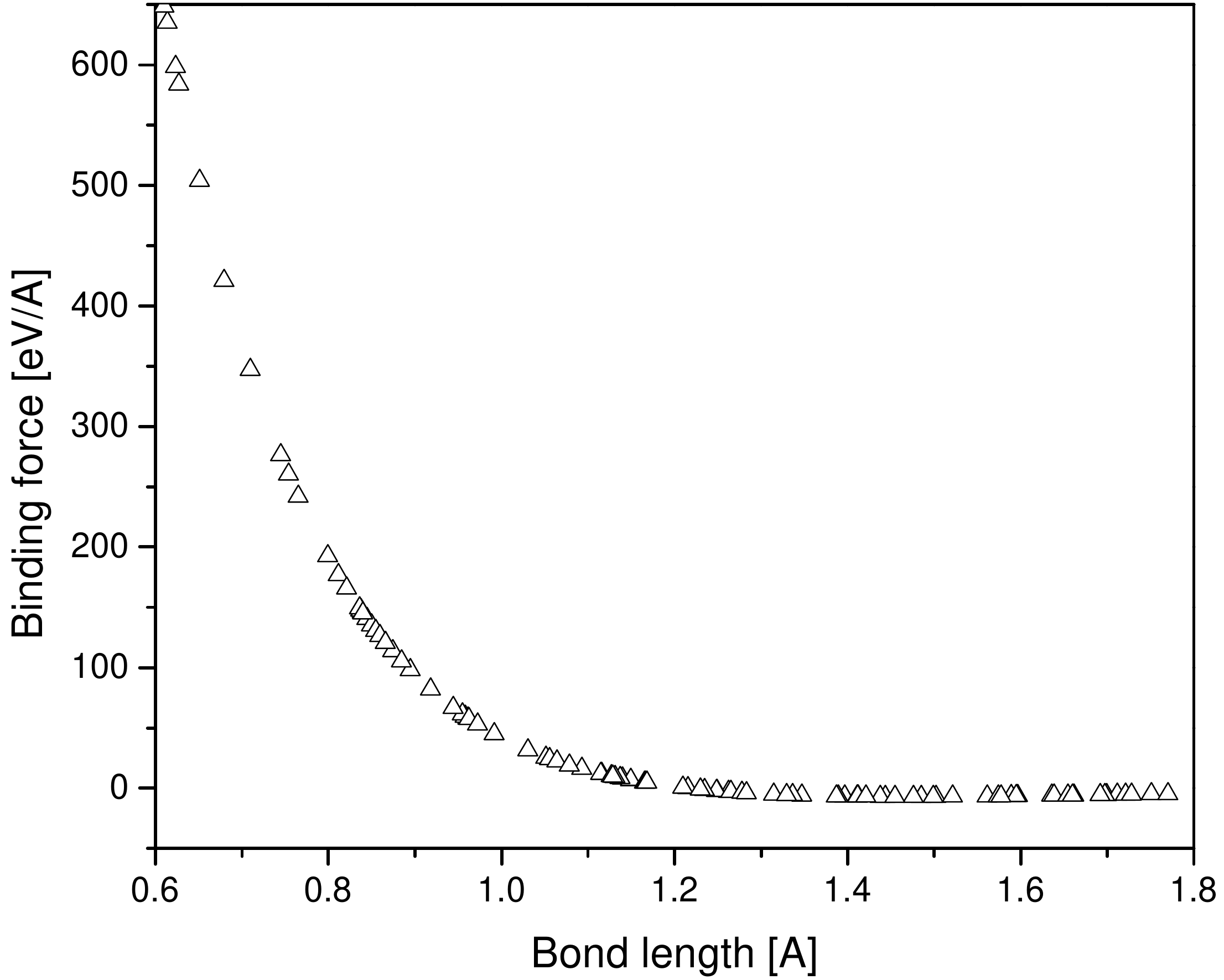}
	\end{subfigure}
	\begin{subfigure}{0.45\textwidth}\caption{} \includegraphics[width=0.99\linewidth]{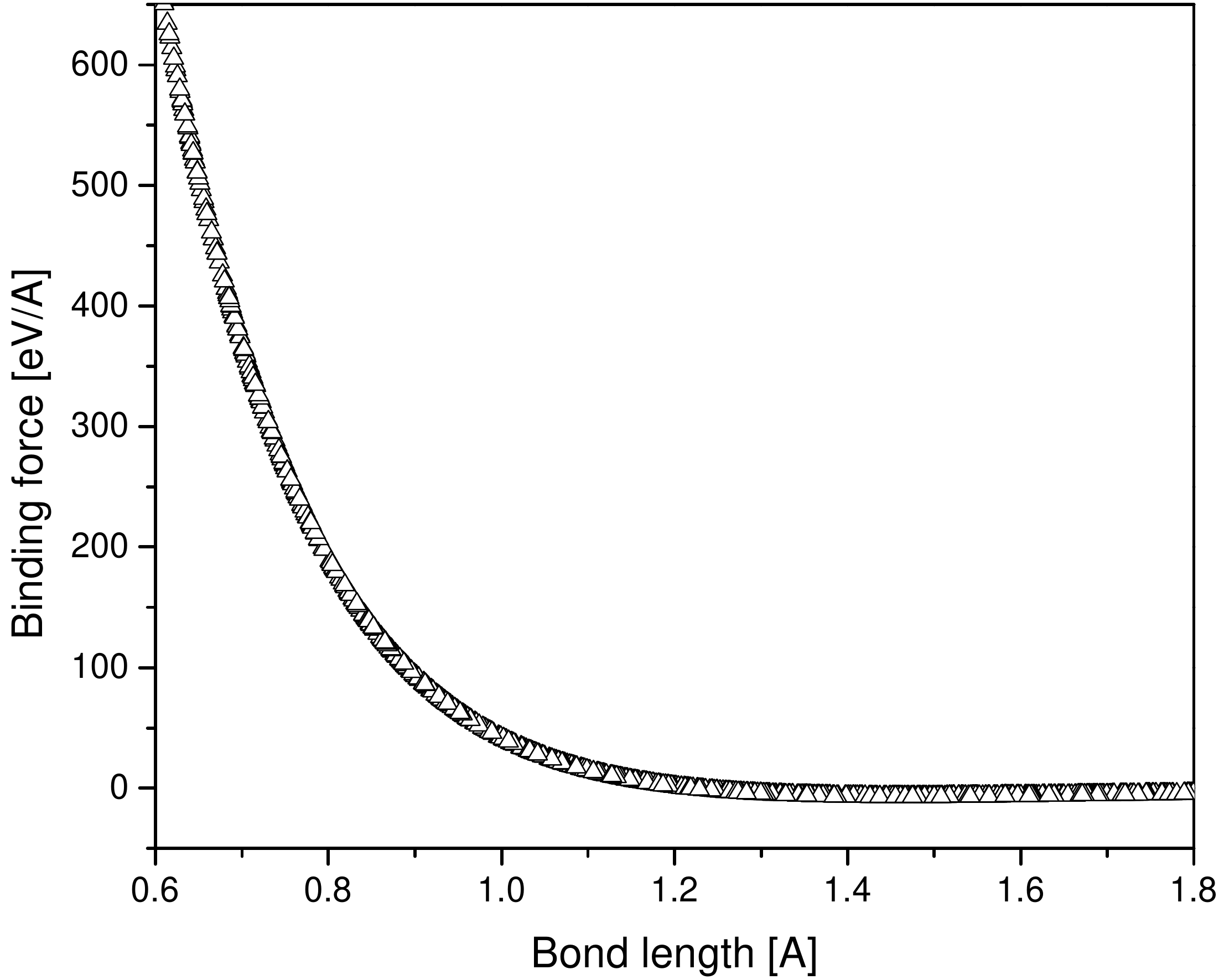}
	\end{subfigure}
	\begin{subfigure}{0.45\textwidth}\caption{} \includegraphics[width=0.99\linewidth]{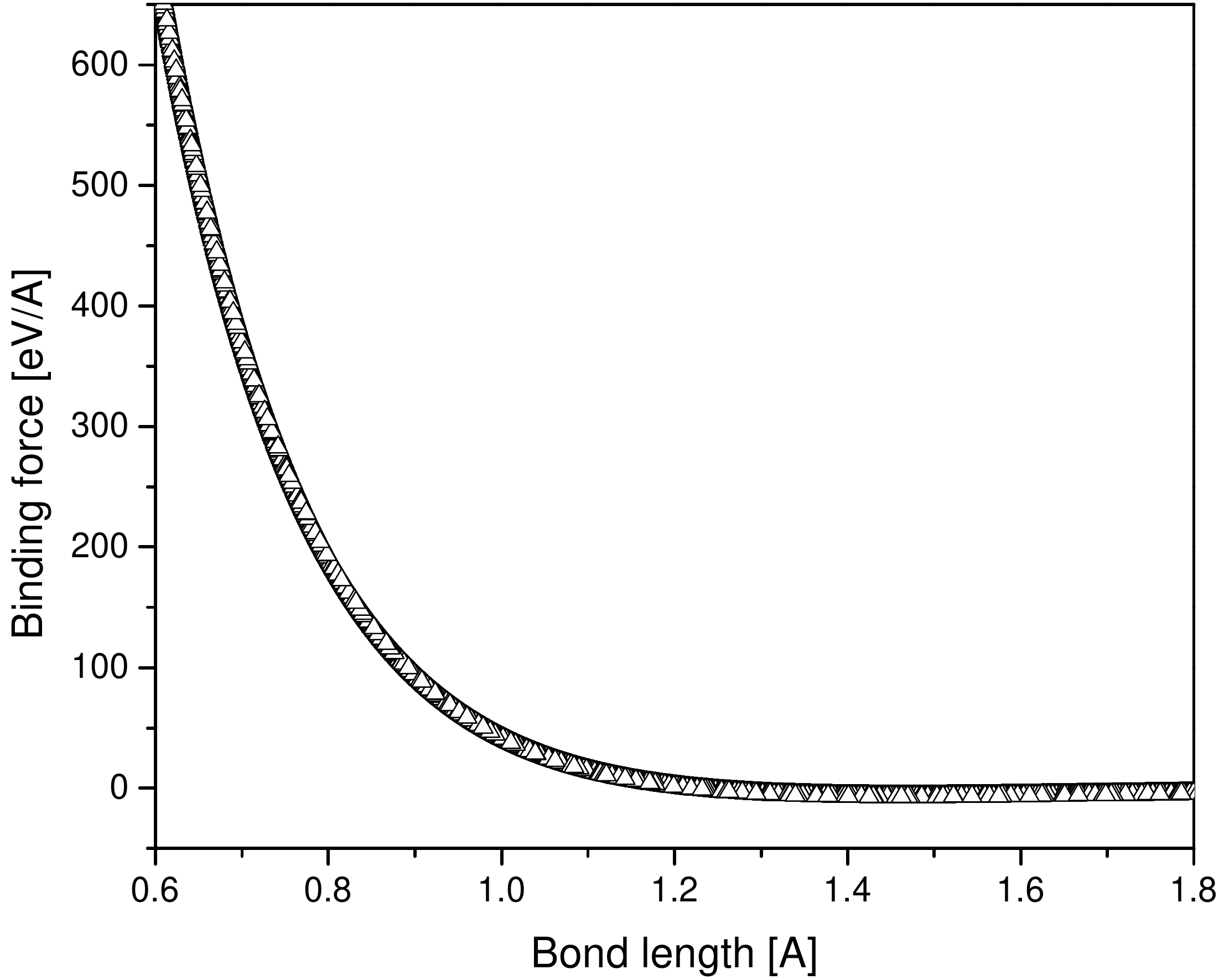}
	\end{subfigure}
	\begin{subfigure}{0.45\textwidth}\caption{} \includegraphics[width=0.99\linewidth]{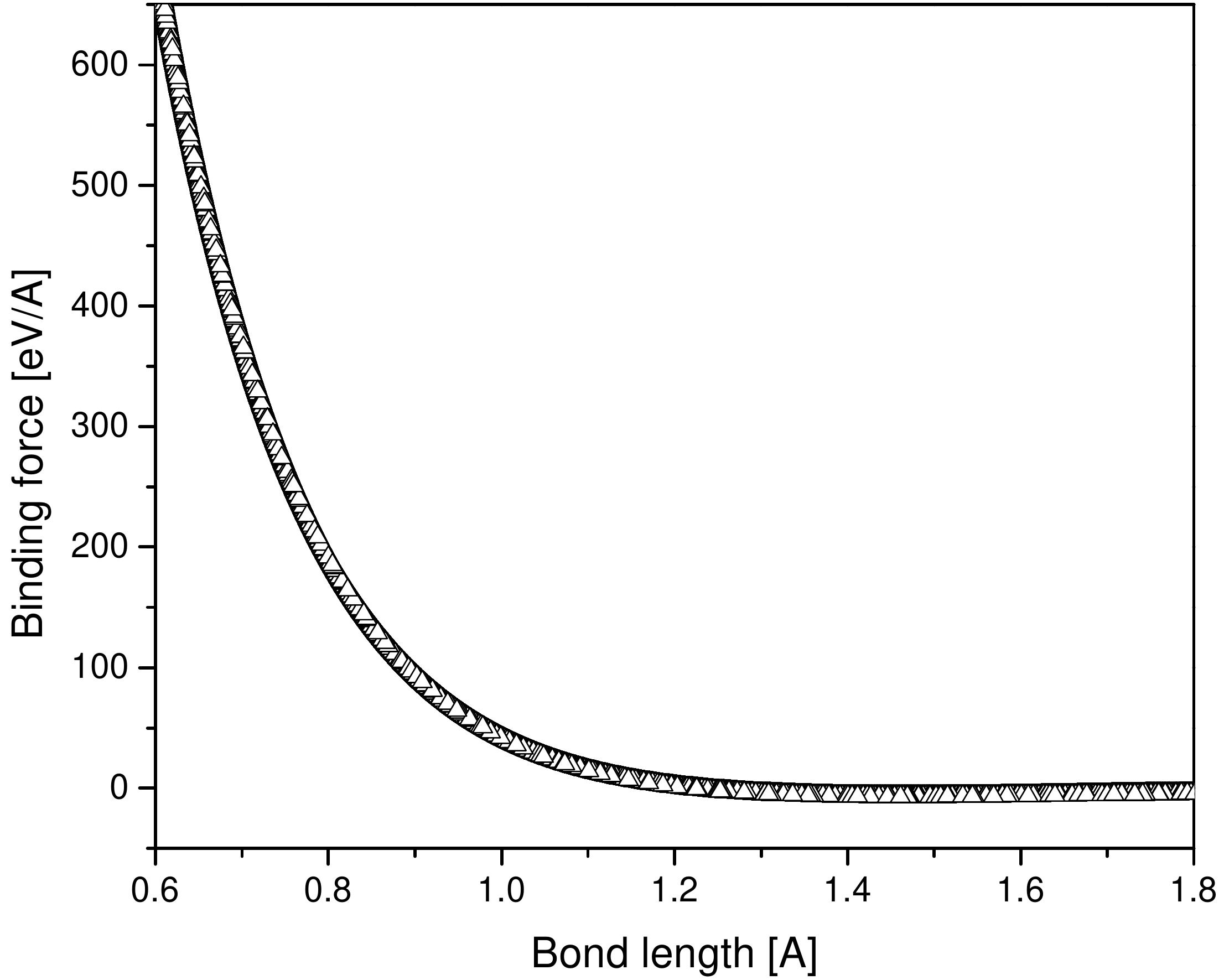}
	\end{subfigure}
    \caption{Molecular oxygen O${}_2$. Sequence of data sets sampled from a Morse potential \cite{Morse:1929} with constants fitted to experimental molecular spectra data \cite{Zielinski:1998, Huber:2022}, cf.~Table~\ref{wp21QK}.. a) $\#D_1 = 100$ points; b) $\#D_2 = 1000$ points; c) $\#D_3 = 10000$ points.; d) $\#D_4 = 100000$ points.} \label{jaI738}
\end{center}
\end{figure}

The computation of the vibrational spectrum of diatomic oxygen ${\rm O}_2$ is an example of a class of problems that arises in chemical physics in connection with the characterization of the molecular spectra of small polyatomic molecules \cite{Herzberg}. Because of the complexity of the quantum many-body problem, the accurate calculation of high-resolution vibrational spectra of even small molecules remains a challenge (cf., e.~g., \cite{Gadzhiev:2013}). The calculations presented here are elementary and intended as a simple numerical example aimed at demonstrating the convergence properties of the DD-Verlet algorithm, Section~\ref{D63kfb}.

\begin{figure}[]
\begin{center}
	\begin{subfigure}{0.8\textwidth}\caption{} \includegraphics[width=0.99\linewidth]{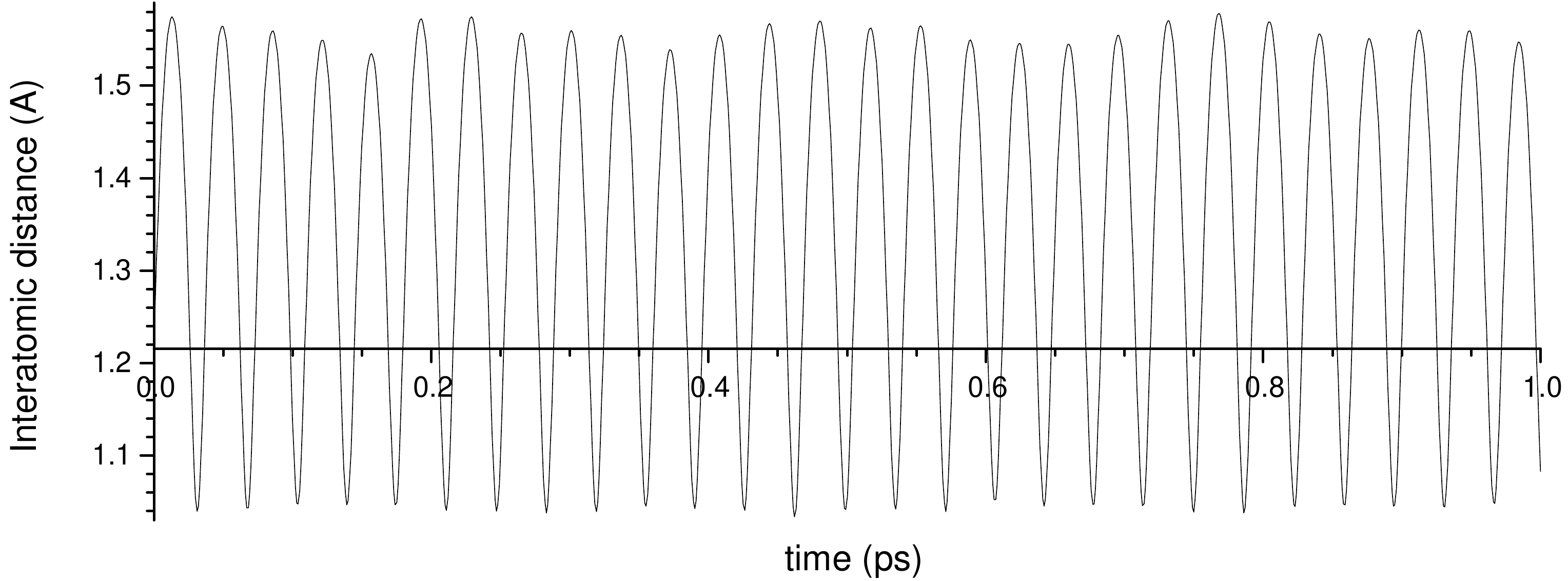}
	\end{subfigure}
	\begin{subfigure}{0.8\textwidth}\caption{} \includegraphics[width=0.99\linewidth]{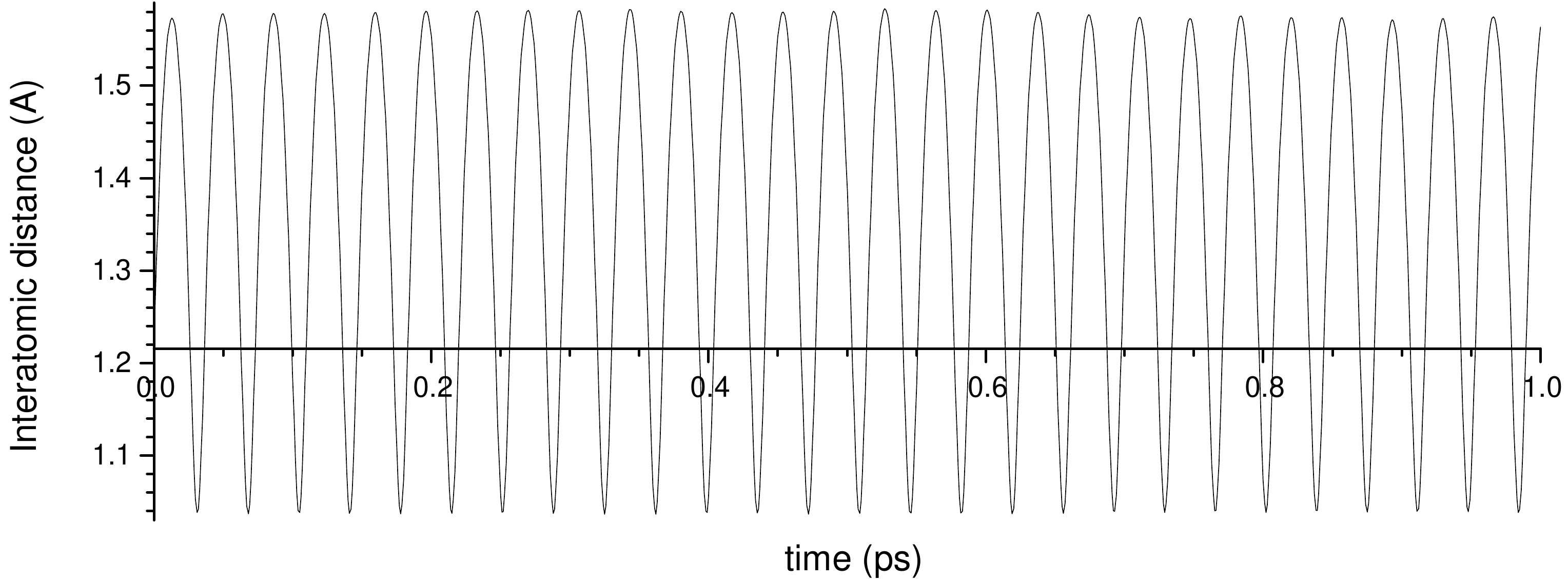}
	\end{subfigure}
	\begin{subfigure}{0.8\textwidth}\caption{} \includegraphics[width=0.99\linewidth]{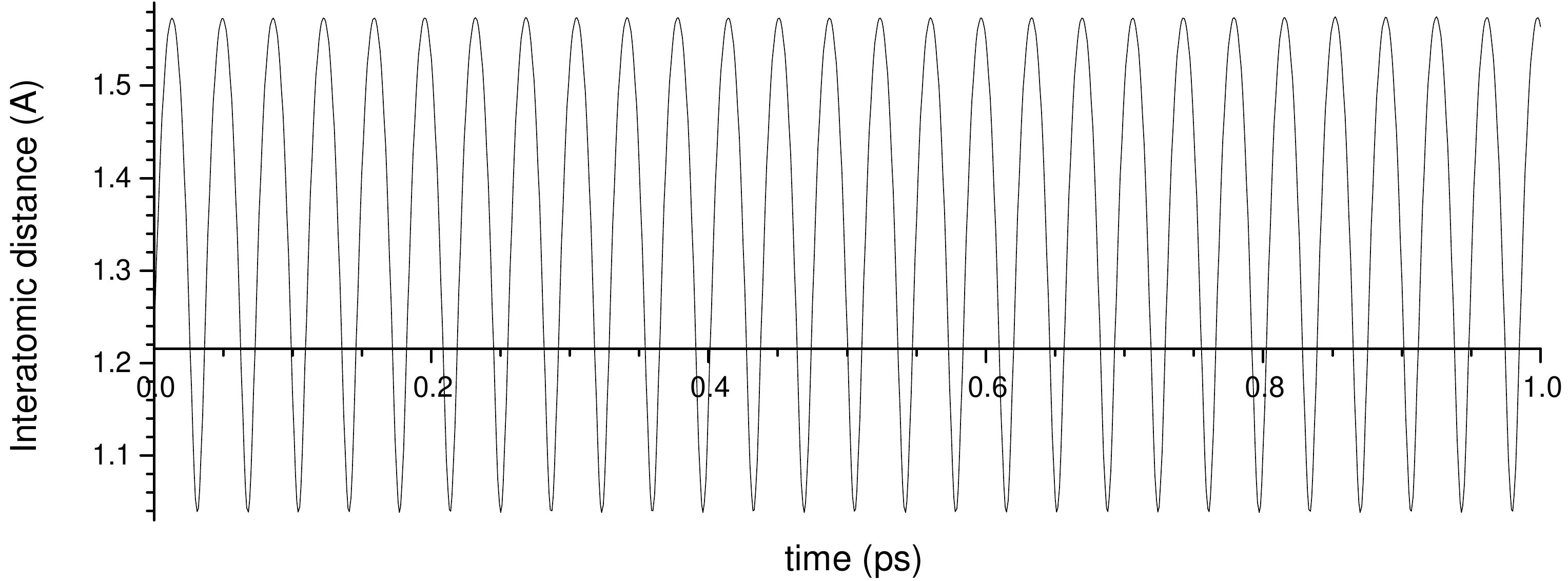}
	\end{subfigure}
    \caption{Molecular oxygen O${}_2$. Sequence of DD solutions computed from the data sets in Fig.~\ref{jaI738}. a) $\#D_1 = 100$ points; b) $\#D_2 = 1000$ points; c) $\#D_3 = 10000$ points.
    } \label{P6o3Qq}
\end{center}
\end{figure}

\setlength{\arrayrulewidth}{0.25mm}
\setlength{\tabcolsep}{18pt}
\renewcommand{\arraystretch}{1.5}

\begin{table}[h!]
\centering
\begin{tabular}{|c|c|c|}
    \hline
    $D_e$ (eV) & $r_e$  (\AA) & $a$ (1/\AA)
    \\  \hline
    5.12931 & 1.21560 & 2.75911
    \\  \hline
\end{tabular}
    \caption{Molecular oxygen O${}_2$. Morse potential constants fitted to experimental molecular spectra data \cite{Huber:2022}. } \label{wp21QK}
\end{table}

For simplicity, we assume that the ${\rm O}_2$ molecule obeys the Morse potential \cite{Morse:1929}
\begin{equation}\label{tGpJ2j}
    V(r)
    =
    D_e
    \Big(
        {\rm e}^{2 a (r_e-r)} - 2 {\rm e}^{a(r_e-r)}
    \Big) .
\end{equation}
with constants fitted to experimental molecular spectra data  \cite{Zielinski:1998, Huber:2022}, cf.~Table~\ref{wp21QK}. The molecule is assumed to undergo anharmonic vibration resulting, e.~g., from a resonance induced by an external magnetic or optical  field. In calculations, the molecule is initially at the equilibrium bond length $r_0=r_e$ and the atoms are imparted an initial relative velocity $v_0 = 50$ \AA/ps. The time step is set to $\tau = 1$ fs throughout.

Synthetic force-field data sets are generated by sampling the Morse force field derived from (\ref{tGpJ2j}) over the range of bond lengths from $0.5 r_e$ to $1.5 r_e$. The resulting force-field data sets are shown in Fig.~\ref{jaI738}. In view of the simplicity of the Morse potential, the agreement with {\sl ab initio} data, Fig.~\ref{atMH3z}, is quite remarkable.

The corresponding DD-Verlet solutions are shown in Fig.~\ref{P6o3Qq}. The $100$-data point solution shows ostensible deviations from the Morse solution. Such deviations notwithstanding, it is remarkable that a qualitatively correct solution, showing the expected smoothness, anharmonicity and periodicity, is obtained from such a small data set. The $1000$-data point and subsequent DD-Verlect solutions are ostensibly converged.

\begin{figure}[ht]
\begin{center}
    \includegraphics[width=0.55\linewidth]{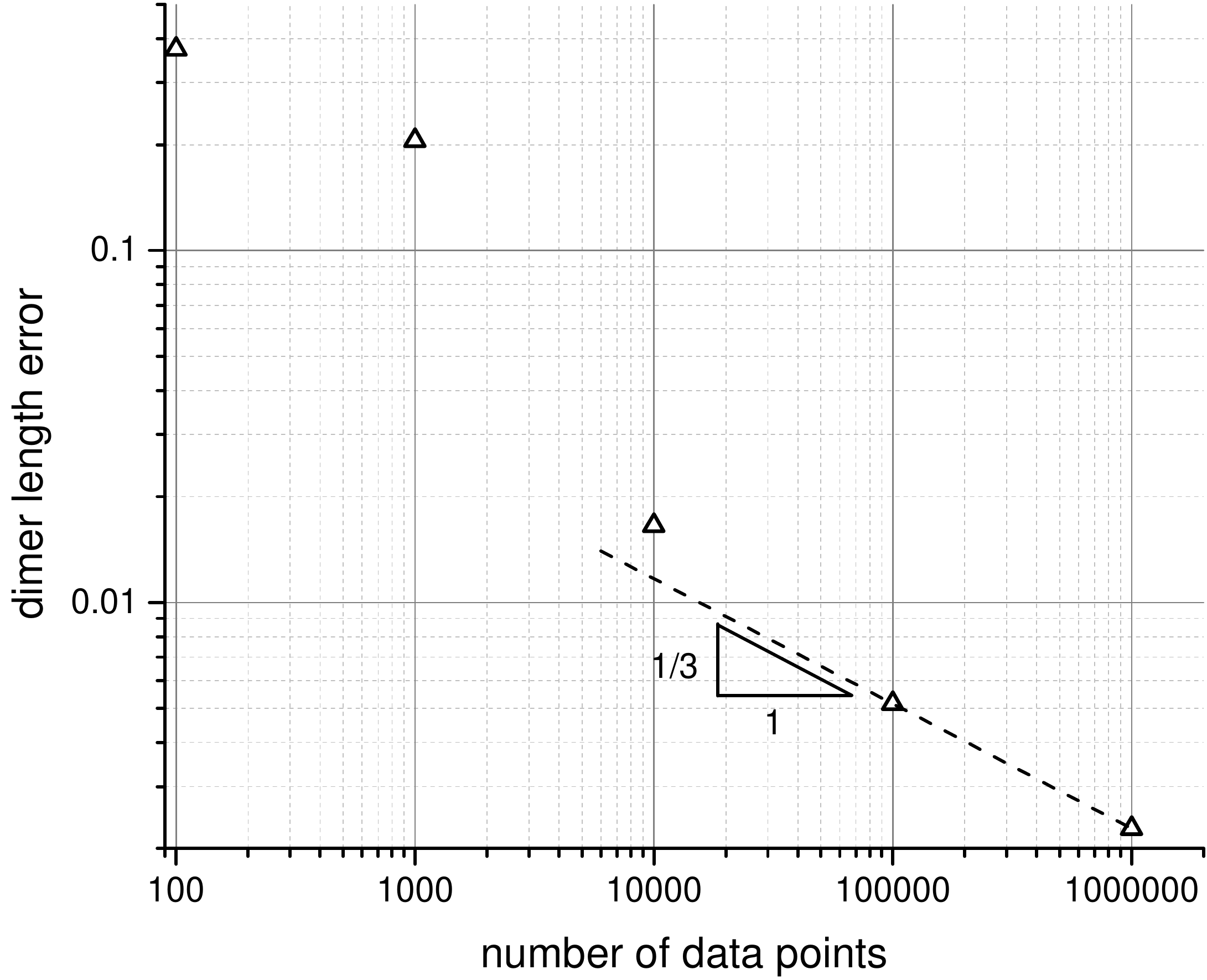}
    \caption{Molecular oxygen O${}_2$. Convergence of DD solutions with respect to the data.} \label{Fjr8rp}
\end{center}
\end{figure}

In order to make the convergence analysis quantitative, we measure the discrepancy between the DD-Verlet and Morse solutions using the weighted $L^2$-norm \cite{Ortiz:1989}
\begin{equation}
    \| u(\cdot) \|
    =
    \Big(
        \int_0^{+\infty}
        | u(t) |^2 \frac{dt}{t^2}
    \Big)^{1/2} .
\end{equation}
The norm quantifies both amplitude and frequency errors in infinite wave-train solutions. The simple identities
\begin{subequations}
\begin{align}
    &
    \| A' \sin\omega t - A'' \sin\omega t\|
    =
    \frac{\pi \omega}{2} | A' - A''|,
    \\ &
    \| A \sin\omega' t - A \sin\omega'' t\|
    =
    \sqrt{\frac{\pi}{2}} | A | \, | \omega' - \omega'' | ,
\end{align}
\end{subequations}
exemplify those properties.

Fig.~\ref{Fjr8rp} shows a log-log plot of the error incurred by the DD-Verlet solutions with respect to the Morse solutions as a function of the size of the force-field data set. As expected from Props.~\ref{bwh2P1} and \ref{dN4PE6}, the DD-Verlet solutions exhibit robust convergence towards the solution of the underlying Morse force field, whence the data is sampled. It bears emphasis that the underlying Morse force-field is assumed unknown for purposes of the DD-Verlet calculations. In addition, all DD-Verlet results stem directly from the sampled force-field data and no intermediate interatomic-potential modeling step is required at any stage of the calculations.

\subsection{Buckminster fullerene ${\rm C}_{60}$}\label{X1H1xV}

We close with an example concerned with the radiation-induced fragmentation of ${\rm C}_{60}$ Buckminster fullerene, or {\sl buckyball} for short. The aim of the example is to demonstrate the ability of the DD-Verlet algorithm to navigate complex conformational transitions using relatively small {\sl ab initio} data sets.

\begin{figure}[ht]
\begin{center}
\begin{subfigure}
    {0.25\textwidth}\caption{} \includegraphics[width=0.99\linewidth]{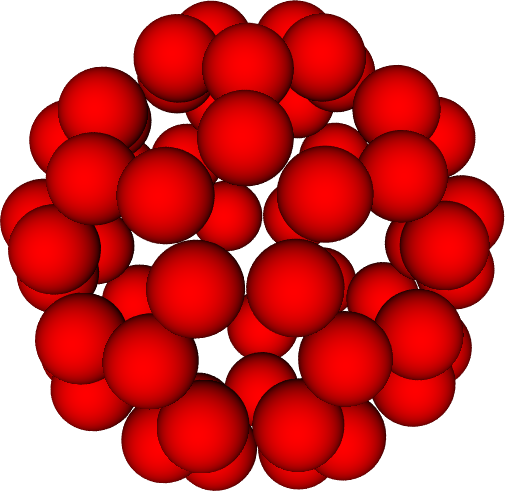}
\end{subfigure}
$\qquad\qquad$
\begin{subfigure}
    {0.25\textwidth}\caption{} \includegraphics[width=0.99\linewidth]{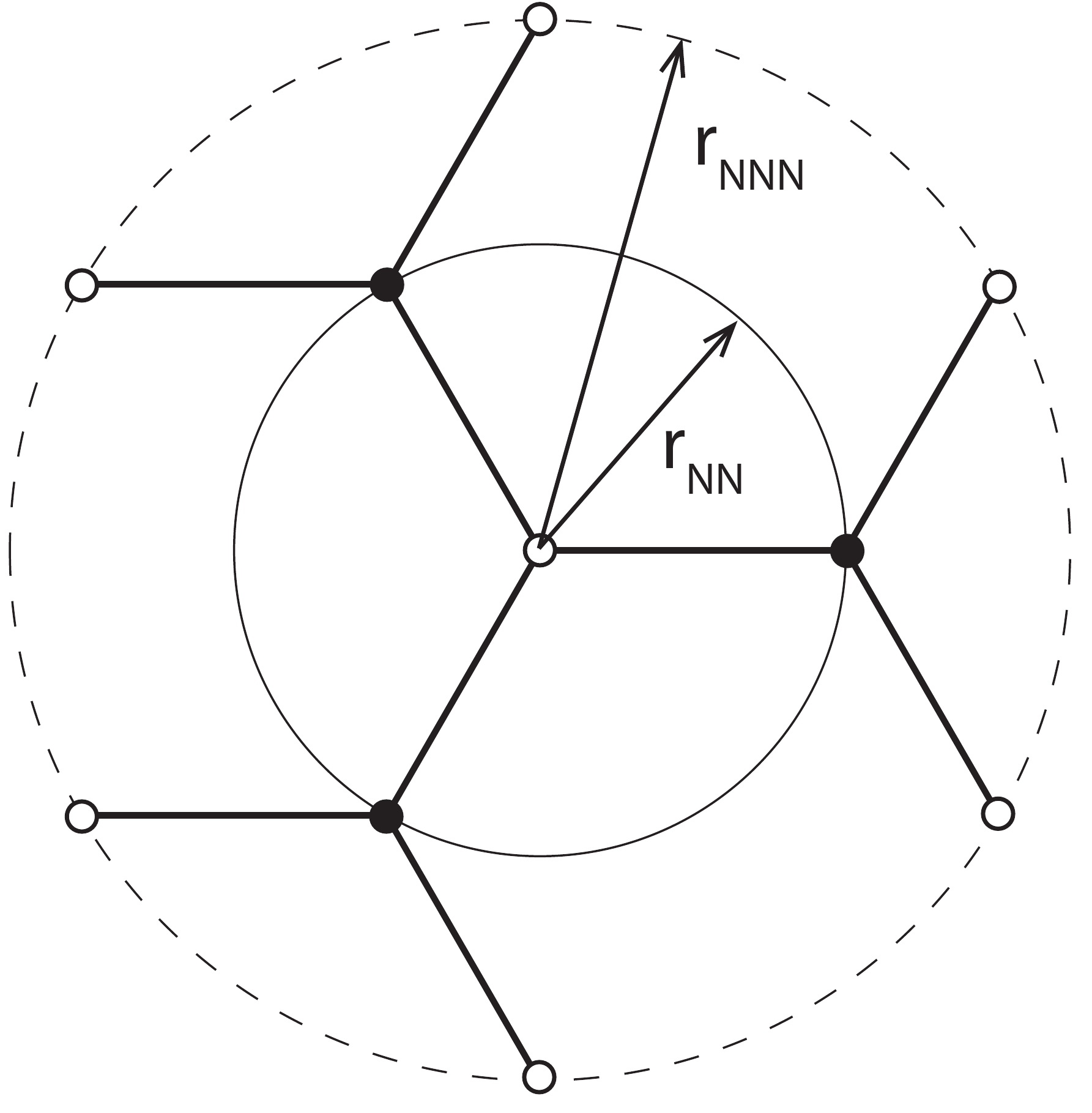}
\end{subfigure}
\caption{a) Relaxed ground-state configuration of the buckminsterfullerene ${\rm C}_{60}$ using the Stillinger-Weber (SW) \cite{SW:1985} potential parametrized as in Table~\ref{1G3iXP}. b) Local environment of an atom in the SW ground-state configuration of ${\rm C}_{60}$ comprising: i) $3$ nearest neighbors (black), at an equilibrium distance $r_{\rm NN} < r_c \equiv$ SW cutoff radius, and ii) $6$ connected next-to-nearest neighbors (white) at a distance $r_{\rm NNN} > r_c$.}\label{22eRFl}
\end{center}
\end{figure}

Buckminster fullerene ${\rm C}_{60}$, discovered in 1985 by Robert Curl, Harold Kroto, and Richard Smalley \cite{Kroto:1985}, is an allotrope of carbon consisting of $60$ carbon atoms in a spatial arrangement of positions and bonds in the form of a dotriacontahedron ($32$-sided figure), Fig.~\ref{22eRFl}a. The term fullerene more generally refers to compounds ${\rm C}_n$ consisting of $n$ carbon atoms, each of which is bonded to three other carbons through two single bonds and one double bond, forming a closed surface.

Interactions of intense ultrashort laser pulses with ${\rm C}_{60}$ fullerene, and the ensuing competition between ionization and fragmentation, have attracted considerable attention \cite{Hertel:2005, kou:2000, Lin:2013} as a means of studying the mechanisms of molecular energy deposition. At low intensity irradiation, the vibrational modes of ${\rm C}_{60}$ are excited, especially its fundamental $A_g$, or {\sl breathing}, mode \cite{Kratschmer:1990, Meilunas:1991}. Beyond a certain intensity, ${\rm C}_{60}$ undergoes {\sl fragmentation} and decomposes into ${\rm C}_n$ fragments \cite{OBrien:1988, Lin:2013}.

For purposes of demonstration, we carry out direct and DD-Verlet calculations based on the Stillinger-Weber (SW) potential \cite{SW:1985}
\begin{equation}
    V_i
    =
    \sum_{j:\, r_{ij}\leq r_{c}} V_2(r_{ij})
    +
    \sum_{k>j:\, r_{ij}\leq r_{c}, \,r_{ik}\leq r_{c}} V_3(r_{ij},r_{ik}) ,
\end{equation}
where the two-body and three-body interaction energies are chosen of the form
\begin{subequations}
\begin{align}
    &
    V_2(r_{ij})
    =
    A
    \Big[B \Big( \frac{|r_{ij}|}{\sigma}\Big)^{-4} - 1 \Big]
    \exp\Big[ \Big( \frac{|r_{ij}|}{\sigma} - b\Big)^{-1}\Big] ,
    \\ \label{5arY5M}
    \begin{split}
        &
        V_3(r_{ij},r_{ik})
        = 
        \lambda
        \exp
        \Big[
        \gamma
        \Big( \frac{|r_{ij}|}{\sigma} - b\Big)^{-1}
        +
        \gamma
        \Big( \frac{|r_{ik}|}{\sigma} - b\Big)^{-1}
        \Big]
        \, \cos^2(\theta_{ijk} - \kappa) ,
    \end{split}
\end{align}
\end{subequations}
with
\begin{equation}
    \theta_{ijk}
    =
    \mathop{\rm acos}
    \frac{r_{ij}\cdot r_{ik}}{|r_{ij}| \, |r_{ik}|}
\end{equation}
representing the angle subtended by the vectors $r_{ij}$ and $r_{ik}$. In calculations, we use the parametrization of \cite{Abraham:1989}, Table~\ref{1G3iXP}.

We note that the SW potential has finite range with cutoff radius $r_c = b \, \sigma = 2.686$\AA $\,$ and an equilibrium bond length $r_e = 1.59169$\AA. From these values, an examination of the local interaction relations (\ref{8MJLzR}) for an atom $i$ in the SW ground-state configuration of ${\rm C}_{60}$  determines the local interaction neighborhood $N_i$ to be of the form shown in Fig.~\ref{22eRFl}b. The local interaction neighborhood comprises: $3$ nearest neighbors (black), at an equilibrium distance $r_{\rm NN} < r_c \equiv$ SW cutoff radius, and ii) $6$ connected next-to-nearest neighbors (white) at a distance $r_{\rm NNN} > r_c$, for a total $\#N_i = 9$. The actual calculation of the Wasserstein distance and the associated transformations is carried using the comparison algorithm of \cite{Bulin:2021}.

\begin{table}[h!]
\centering
\begin{tabular}{|c|c|c|c|c|c|c|}
    \hline
    $A$ (eV) & $B$ & $b$ & $\lambda$ (eV) & $\gamma$ & $\kappa$ & $\sigma$ (\AA)
    \\  \hline
    5.3790 & 0.5082 & 1.8945 & 18.7079 & 1.2 & -0.5 & 1.41800
    \\  \hline
\end{tabular}
    \caption{Stillinger-Weber (SW) potential parameters for carbon \cite{Abraham:1989}.} \label{1G3iXP}
\end{table}

\begin{figure}
\begin{center}
\begin{subfigure}
    {0.9\textwidth}\caption{} \includegraphics[width=0.99\linewidth]{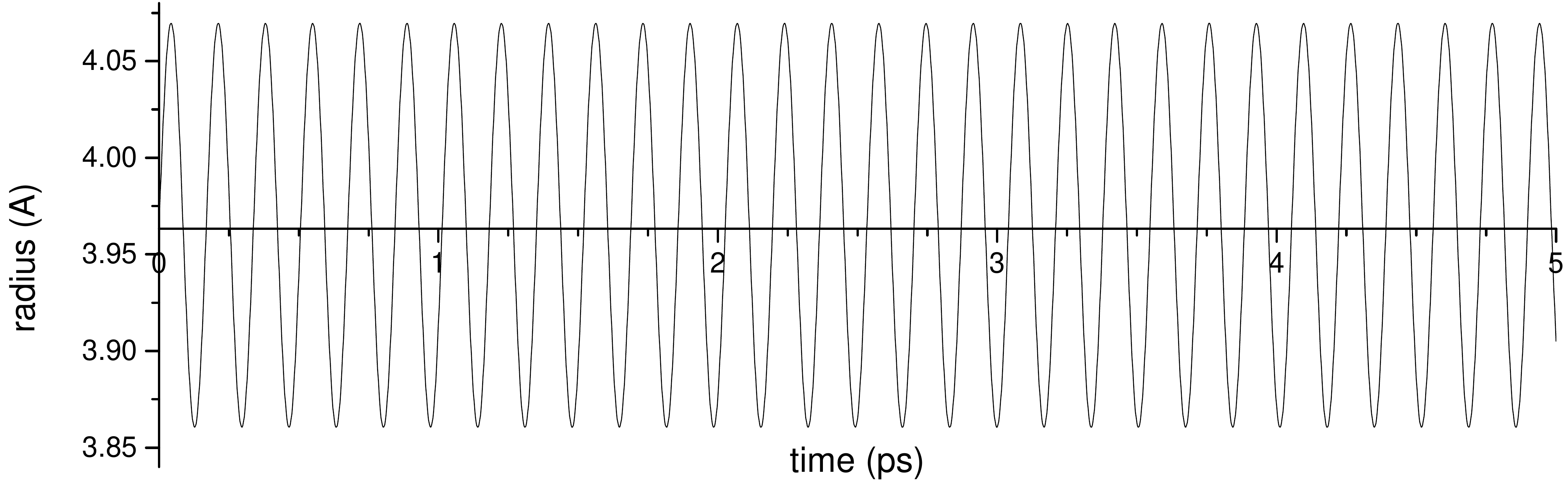}
\end{subfigure}
\vglue 0.5truecm
\begin{subfigure}
    {0.56\textwidth}\caption{} \includegraphics[width=0.99\linewidth]{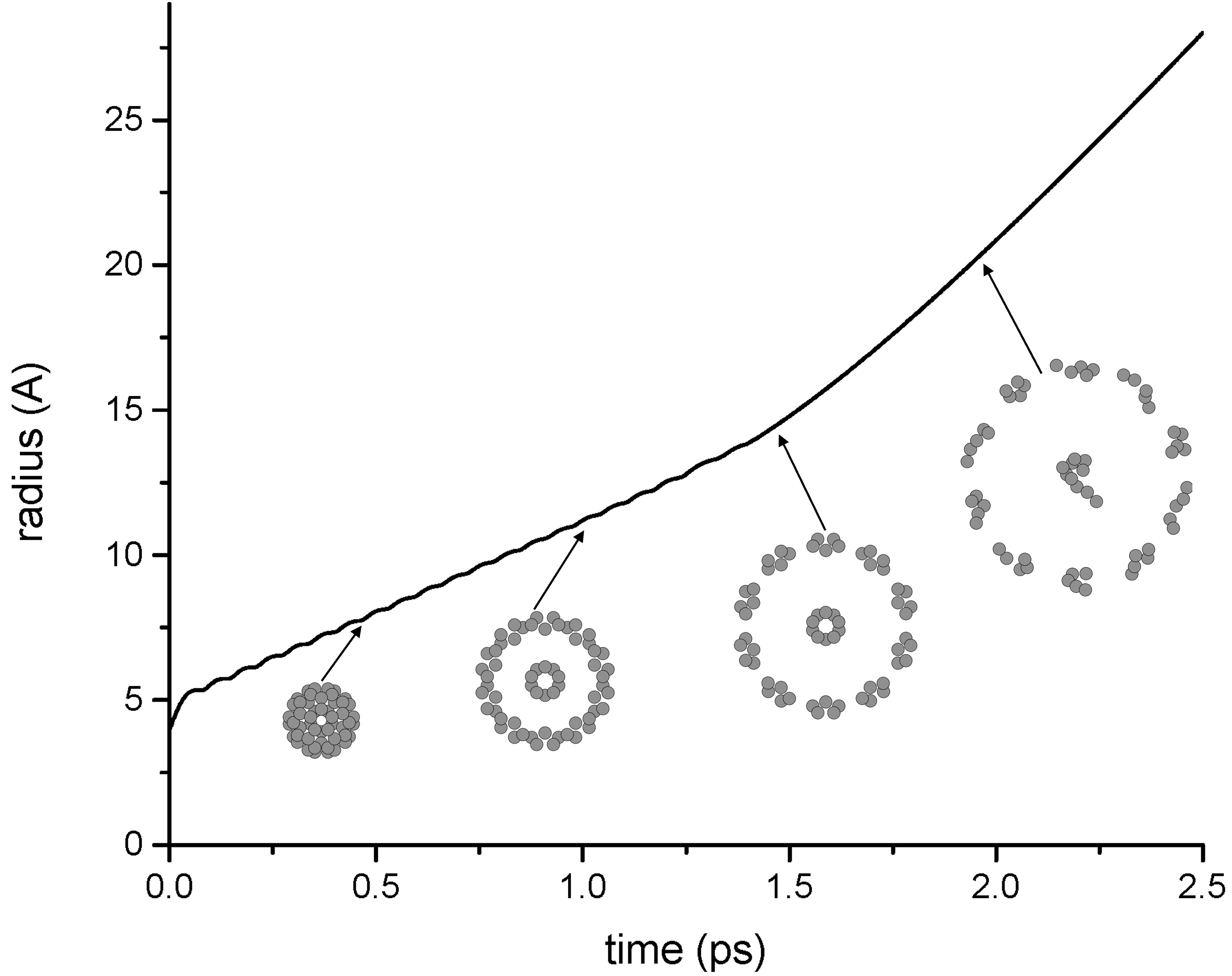}
\end{subfigure}
\caption{Buckminsterfullerene ${\rm C}_{60}$. Time evolution of the average radius predicted by the SW potential after: a) low-intensity irradiation; b) high-intensity irradiation. Inset: computed configurations at $0.5$ps intervals (plotted to scale).} \label{F4zpbe}
\end{center}
\end{figure}

Fig.~\ref{F4zpbe} shows the time evolution of the average radius of the ${\rm C}_{60}$ molecule after irradiation as computed directly from the SW potential. Fig.~\ref{F4zpbe}a corresponds to an initial radial velocity $v_0 = 0.00382$\AA/fs, below the fragmentation threshold, and Fig.~\ref{F4zpbe}b to an initial radial velocity $v_0 = 0.0382$\AA/fs, above the fragmentation threshold. In the former case, the fullerene undergoes an anharmonic vibration in the breathing mode, whereas in the latter case the molecule initially dissociates into identical pentarings ${\rm C}_5$ that scatter ballistically. A sequence of snapshots showing the scattering of the pentarings following fragmentation is shown inset in Fig.~\ref{F4zpbe}b.

\begin{figure}
\begin{center}
	\begin{subfigure}
        {0.30\textwidth}\caption{} \includegraphics[width=0.99\linewidth]{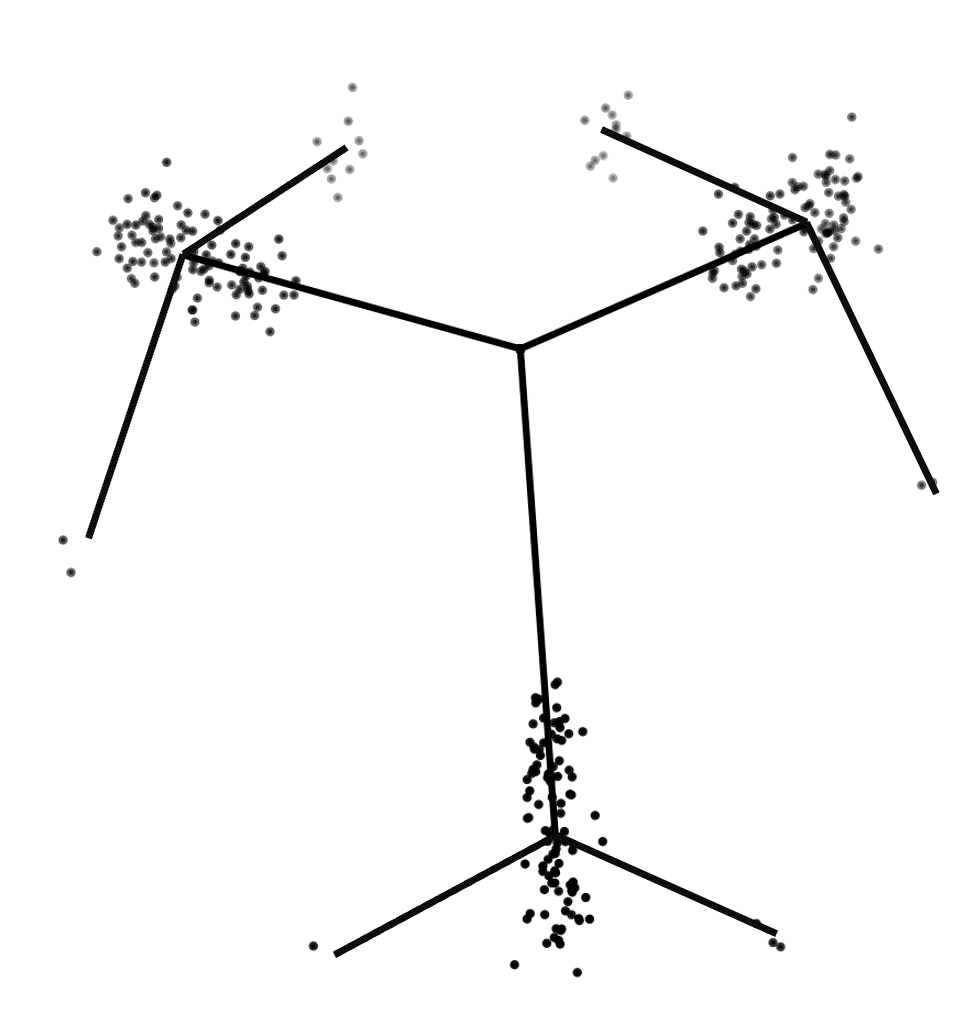}
	\end{subfigure}
	\begin{subfigure}
        {0.30\textwidth}\caption{} \includegraphics[width=0.99\linewidth]{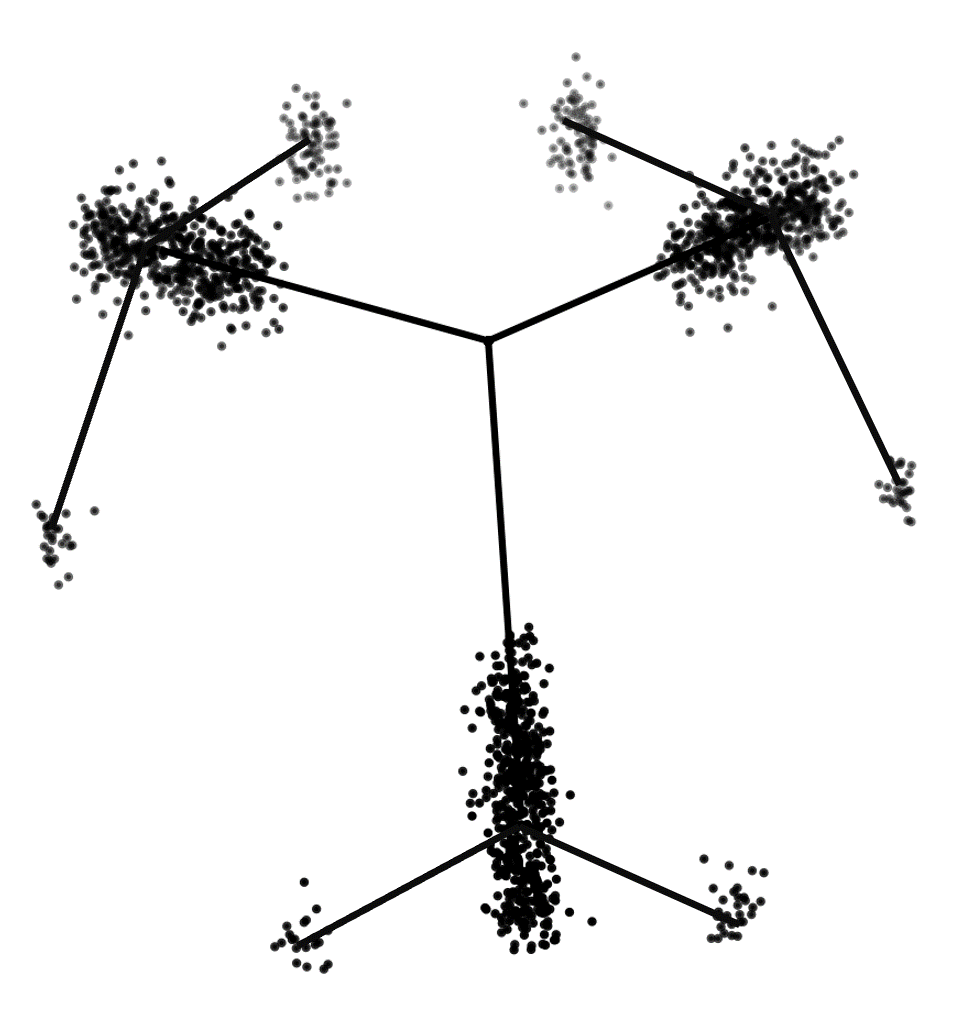}
	\end{subfigure}
	\begin{subfigure}
        {0.30\textwidth}\caption{} \includegraphics[width=0.99\linewidth]{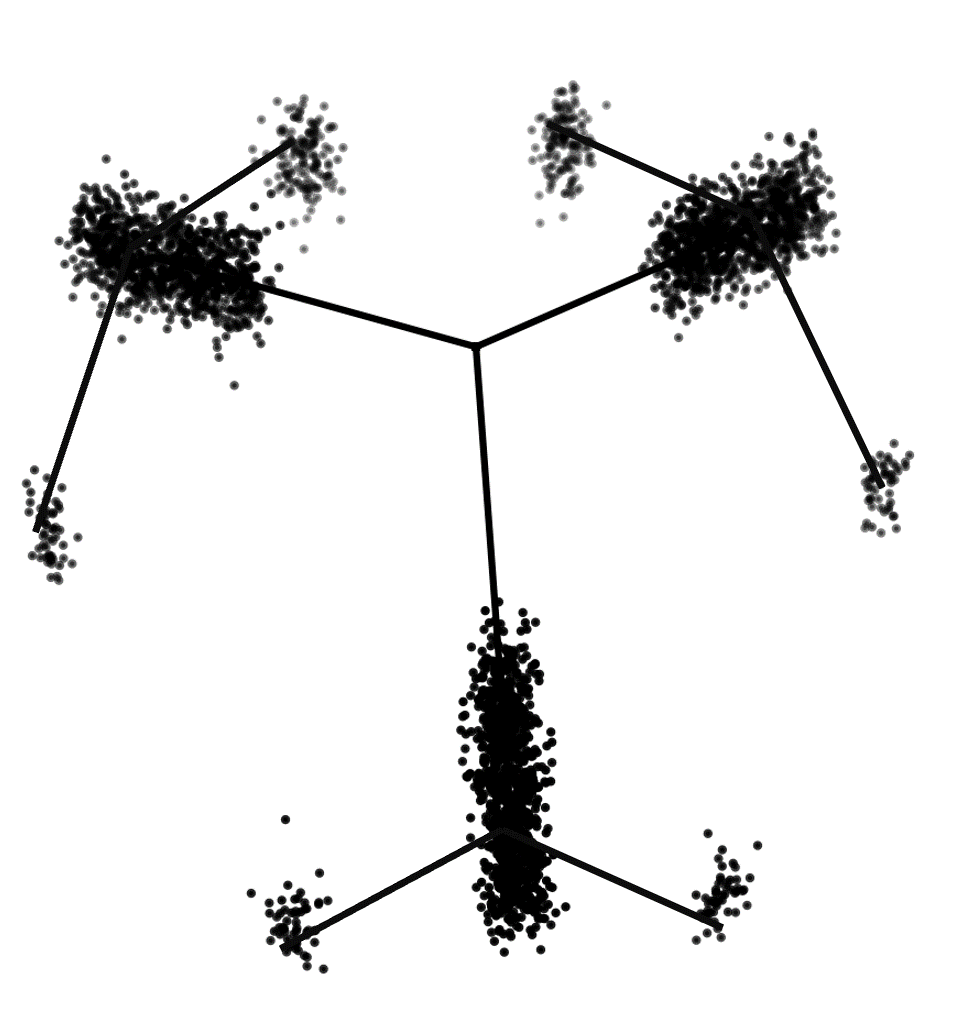}
	\end{subfigure}
    \caption{Local configuration data sets for assumed cutoff radius $r_c=2$\AA. Number of points: a) $\#D_{\rm loc} = 100$; b) $\#D_{\rm loc} = 500$; c) $\#D_{\rm loc} = 1000$.} \label{b0yjFV}
\end{center}
\end{figure}

We wish to ascertain whether these behaviors can be reproduced by DD-Verlet calculations based on force-field data sets sampled from the SW potential. To this end, we generate synthetic force-field data sets that tabulate local interaction neighborhoods $N_i$ and corresponding forces on the central atom $r_i$. Since the dimensionality is too large for local configurations to be sampled uniformly, we adopt a {\sl goal-oriented}, or {\sl importance-sampling}, sampling strategy, whereby the goal is to bias the sampling towards local configurations that are likely to arise under the specific conditions of the application of interest. Since the trajectories followed by the ${\rm C}_{60}$ molecule after low-intensity irradiation are expected to consist predominantly of radial expansion/contraction, we generate local configurations  by randomly expanding/contracting the ground-state configuration, relaxing it and adding random noise to the positions of the atoms. Since the effective cutoff radius $r_c$, if any, is presumed unknown {\sl a priori}, we generate data sets over a range of assumed values of $r_c$.  Examples of local configuration sets generated by this procedure assuming a cutoff radius $r_c = 2$\AA $\,$ are shown in Fig.~\ref{b0yjFV}.


\begin{figure}
\begin{center}
\includegraphics[width=0.85\linewidth]{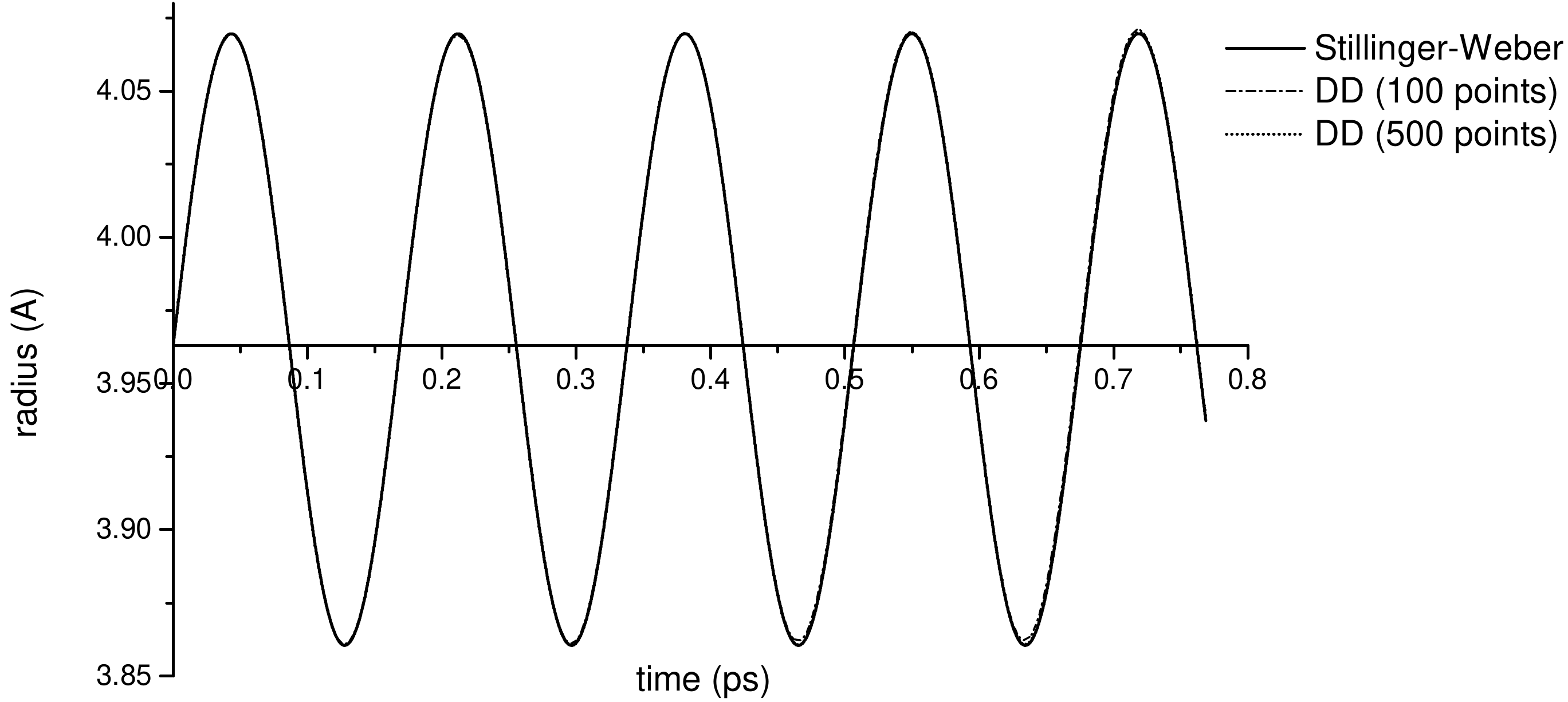}
    \caption{DD-Verlet simulation of ${\rm C}_{60}$ ($r_c=2$\AA). Time evolution of the average radius after low-intensity irradiation. Comparison of trajectories obtained directly from the SW potential and from DD-Verlet for local force-field data sets of size $100$ and $500$.} \label{fig:convergence_2}
\end{center}
\end{figure}

Fig.~\ref{fig:convergence_2} shows results of low-intensity irradiation DD-Verlet calculations for data sets generated assuming $r_c=2$\AA $\,$ compared to the direct calculations. As may be seen from the figure, the DD-Verlet trajectories do closely match the results of the direct SW calculation even for the small data set of size $100$.

\begin{figure}
\begin{center}
	\begin{subfigure}
        {0.475\textwidth}\caption{} \includegraphics[width=0.99\linewidth]{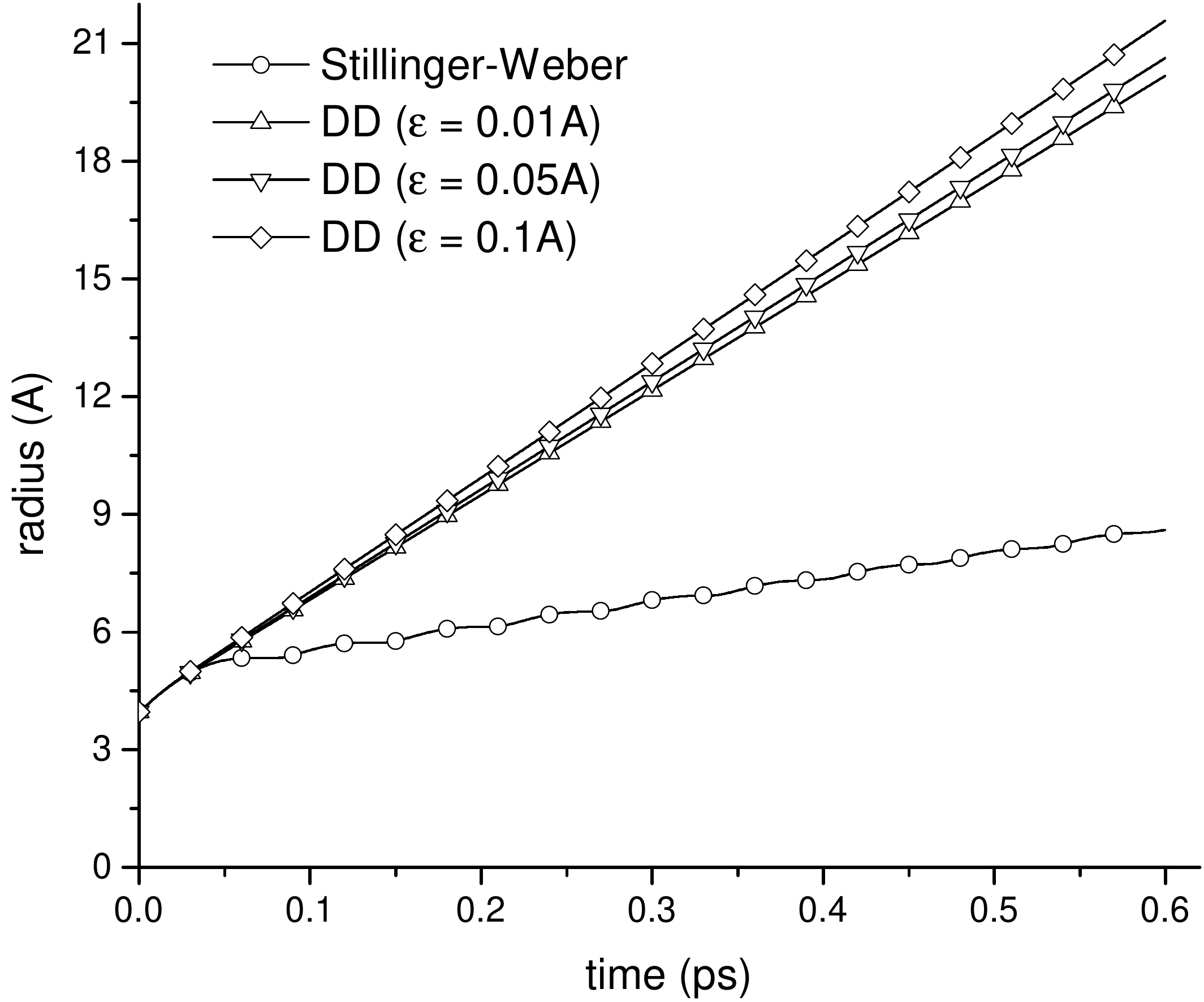}
	\end{subfigure}
	\begin{subfigure}
        {0.475\textwidth}\caption{} \includegraphics[width=0.99\linewidth]{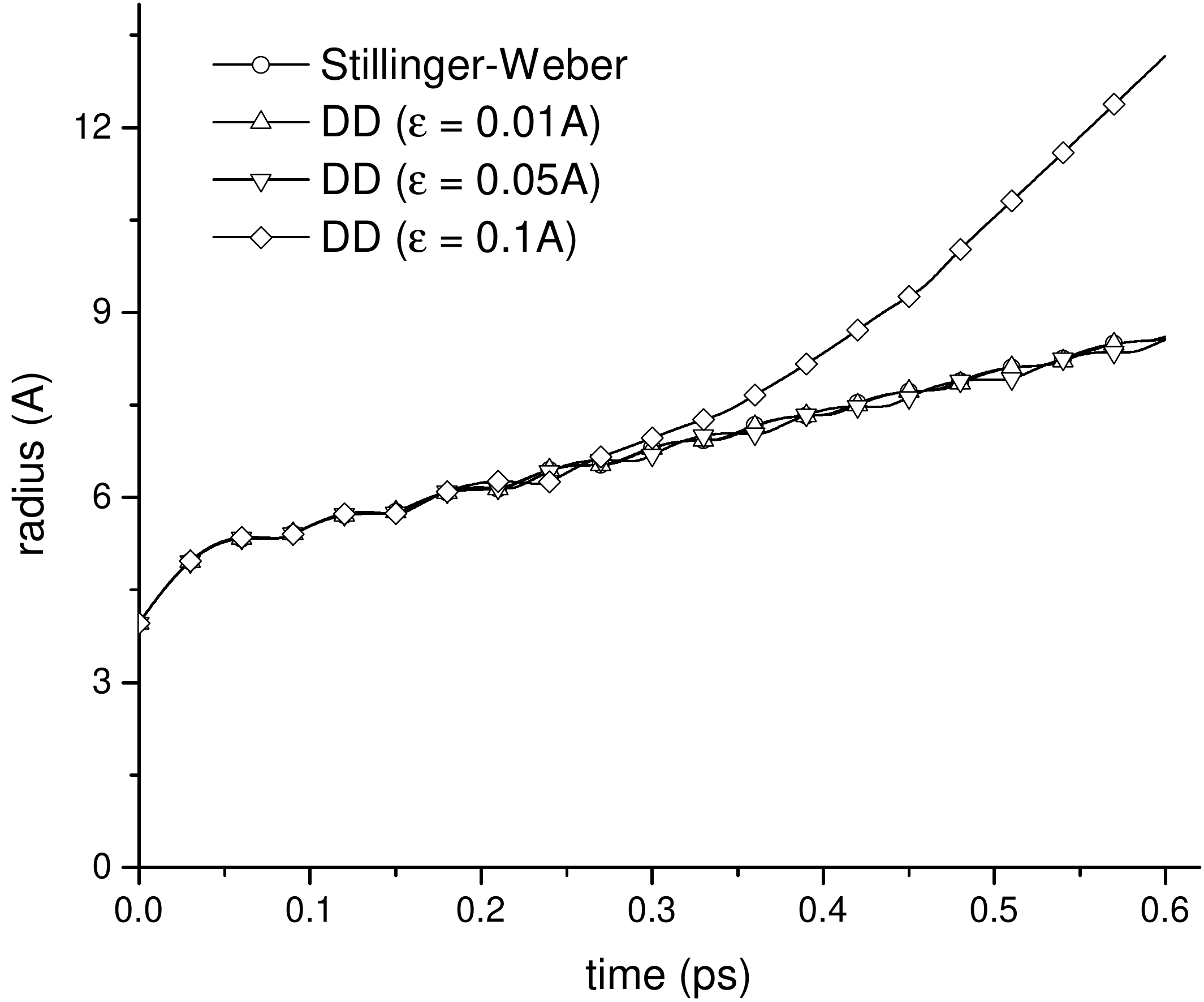}
	\end{subfigure}
    \caption{DD-Verlet simulation of ${\rm C}_{60}$. Time evolution of the average radius after high-intensity irradiation. Comparison of trajectories obtained directly from the SW potential and from DD-Verlet with sampling tolerance $\epsilon = 0.1$\AA, $0.05$\AA and $0.01$\AA. a) Local force-field data sampled using a cutoff radius $r_c = 2$\AA; b) Local force-field data sampled using a cutoff radius $r_c = 4$\AA.} \label{4XNfoa}
\end{center}
\end{figure}

Generating good force-field data sets for the high-intensity irradiation case {\sl a priori} is challenging, as many different structural changes must be covered by the data in order to reproduce the fragmentation behavior of the fullerene. We may circumvent this difficulty by exploiting the variational structure of the DD-Verlet solver in order to adapt the force-field data sets {\sl on-the-fly}, in the spirit of {\sl active learning} \cite{Settles:2012}. Thus, we recall that DD-Verlet uses the Wasserstein distance ${\rm dist}(r,s)$ from a local configuration $r$ to find similar local configurations $s$ in the force-field data set. Therefore, situations in which the trajectory wanders into poorly sampled regions of phase space are immediately identified, since they result in large distances to the data. To correct for such sampling deficiencies, we perform an explicit Stillinger-Weber calculation whenever in the calculation we encounter a local configuration $r$ such that no local configuration $s$ in the force-field data set satisfies the condition ${\rm dist}(r,s)\leq\epsilon$, for some prespecified tolerance $\epsilon$, i.~e., if the distance from $r$ to the data set exceeds the tolerance.

\begin{table}[h!]
\centering
\begin{tabular}{|c|c|c|c||}
    \hline
    Stillinger-Weber & DD ($\epsilon=0.1$\AA) & DD ($\epsilon=0.05$\AA) & DD ($\epsilon=0.01$\AA)
    \\  \hline
    6000 & 269 & 370 & 1460
    \\  \hline
\end{tabular}
    \caption{DD-Verlet simulation of ${\rm C}_{60}$. Response up to $0.6$ps following high-intensity irradiation. Number of force-evaluations for different values of the distance tolerance $\epsilon$.}
    \label{table:force_count}
\end{table}

As may be seen from Fig.~\ref{4XNfoa}a, for a neighborhood cutoff radius of $r_c = 2$\AA $\,$, DD-Verlet exhibits robust convergence with respect to the data-set tolerance (and thus the data-set size) but converges to a trajectory that is at variance with SW. Evidently, in hindsight this behavior is to be expected since the data is generated assuming a cutoff radius that underestimates the actual value $r_c = 2.686$\AA. This underestimation results in local force-field data sets with poor coverage of the actual SW force-field. By contrast, Fig.~\ref{4XNfoa}b collects similar curves obtained from force-field data sets generated assuming a cutoff radius $r_c = 4$\AA. As may be seem from the figure, in this case the DD-Verlet trajectories converge robustly to the correct SW trajectory as the sampling tolerance $\epsilon$ is decreased. Remarkably, convergence is achieved for modest force-field data set sizes, Table~\ref{table:force_count}.

\section{Concluding remarks}

We have presented a paradigm that enables molecular dynamics calculations to be performed directly from sample data, e.~g., obtained from {\sl ab initio} calculations, thereby eschewing the conventional modeling of the data by empirical interatomic potentials entirely. The DD paradigm is {\sl lossless}, i.~e., it uses the available data in its entirety; {\sl unbiased} in the sense of being free of modeling assumptions, {\sl ad hoc} representational 'features', parametrization and regression choices, and other extraneous elements; and, by its direct and exclusive reliance on data, it is {\sl modeling-error free}.

The data used by the DD paradigm specifically samples local configurations and corresponding atomic forces and is, therefore, {\sl fundamental}, i.~e., it is not beholden to any particular model. This strict reliance on fundamental data is potentially far-reaching as it facilitates data generation, management and sharing. Thus, fundamental data is by its very nature {\sl fungible}, i.~e., data sets of diverse provenance can be {\sl merged} directly to obtain larger data sets. This property opens the way to the standing of publicably-editable, public-domain, force-field data sets that can be built collaboratively worldwide. Considerations of sampling error, reliability and data provenance can easily be incorporated into the DD paradigm as quality-control measures \cite{Kirchdoerfer:2017}.

The computation of distances between clusters of atoms and the need to execute fast searches in large data sets are among the main computational bottlenecks of the DD paradigm. For large local atomic clusters, it becomes impractical to compute the discrete Wasserstein distance by examining exhaustively all the rearrangements of the atoms, which is an operation of combinatorial complexity. When the sampling is sufficiently dense, the ideas in \cite{Bulin:2021} can be used to limit the combinatorial complexity of the distance calculation to cubic order. In other cases, a Kantorovich relaxation, which reduces the operation to the solution of a linear programming problem, interior-point regularizations thereof, and generally considerations of efficiency in the computation of the distance become of the essence. In addition, fast search algorithms for purposes of Big Data analysis have been extensively investigated and enable the efficient utilization of billion-size data sets (cf., e.~g., \cite{Eggersmann:2021} and references therein).

Evidently, the local force-field data sets used in the DD calculations need to provide adequate {\sl coverage} of the regions of phase space traversed by the trajectories of interest. However, data sets covering the entirety of phase space uniformly are not possible and the sampling must of necessity be {\sl goal-oriented}, i.~e. suited to a particular class of trajectories. For instance, in the application to irradiated ${\rm C}_{60}$ fullerenes presented above, the expectation that the molecules undergo predominantly radial expansion/contraction at low irradiation intensities has been used in order to fashion {\sl a priori} goal-oriented data sets delivering fast convergence.

In more general settings, it is not always obvious at the outset whether a given data set is likely to supply adequate coverage. Conveniently, the DD paradigm is {\sl self-correcting} in that regard. Thus, a trajectory that wanders into a poorly sampled region of phase space inevitably results in large distances to the data set being recorded during the calculations, which immediately flags the data as insufficient for the intended purpose. Furthermore, the poorly sampled regions of phase space thus identified can then be sampled by carrying out additional {\sl ab initio} calculations, with the process repeated until the distance between the computed trajectory and the data set is below a prespecified tolerance at all times. The resulting iteration, something referred to to as {\sl active learning} \cite{Settles:2012}, sets forth an approximating scheme with respect to the data itself. If convergent, the scheme results in trajectories that are close to those of the underlying--and unknown--force field sampled by the data, a remarkable feat indeed.

These and other questions and improvements of the DD paradigm presented here suggest themselves as worthwhile directions for further research.

\section*{Acknowledgments}

This work has been funded by the Deutsche Forschungsgemeinschaft (DFG, German Research Foundation) {\sl via} project 211504053 - SFB 1060; project 441211072 - SPP 2256; and project 390685813 -  GZ 2047/1 - HCM. Hausdorff Center for Mathematics (HCM). P.~Ariza gratefully acknowledges financial support from the Consejer\'ia de Econom\'ia y Conocimiento of the Junta de Andaluc\'ia, Spain, under grant number P18-RT-1485; and from Ministerio de Ciencia, Innovación y Universidades
under grant number RTI2018-094325-B-I00.

\begin{appendix}
\section{Proofs of Propositions \ref{bwh2P1} and \ref{dN4PE6}}

\begin{proof} (of Prop.~\ref{bwh2P1})
For simplicity, we assume that all atoms have identical mass $m_i = m$. We work directly with the Euler-Lagrange equations (\ref{qhw8ZY}). Integrating (\ref{Q1Z9bs}) twice using (\ref{5pCIel}) we obtain
\begin{equation}
    w_h(t) = - \int_0^t\Big( \int_0^{t'} \frac{1}{m} ( r_h(t'') - s_h(t'') ) \, dt'' \Big) \, dt' .
\end{equation}
Estimating,
\begin{equation}\label{mGDD0s}
    | w_h(t) |
    \leq
    \int_0^t\Big( \int_0^{t'} \frac{1}{m} | r_h(t'') - s_h(t'') | \, dt'' \Big) \, dt' .
\end{equation}
By the assumption on the density of sampling, we have $| r_h(t) - s_h(t) | \leq \epsilon_h$, whereupon (\ref{mGDD0s}) gives
\begin{equation}\label{tLG23q}
    |w_h(t)| \leq \frac{T^2}{m} \epsilon_h ,
\end{equation}
which shows that the Lagrange multiplier $w_h(\cdot)$ tends to zero uniformly over $[0,T]$. Let $r(\cdot)$ be the exact solution of the initial-value problem (\ref{5xvxXr}). Subtracting equations of motion and initial conditions, we obtain
\begin{subequations}
\begin{align}
    &
    m (\ddot{r}_h(t)-\ddot{r}(t)) + g_h(t) - f(r(t)) = \kappa^{-2} w_h(t) .
    \\ & \label{8qRvRa}
    r(0) - r_h(0) = 0,
    \quad
    \dot{r}(0) - \dot{r}_h(0) = 0 .
\end{align}
\end{subequations}
Integrating with respect to time and using the second of (\ref{8qRvRa}),
\begin{equation}
    m (\dot{r}_h(t)-\dot{r}(t))
    =
    \int_0^t \big( \kappa^{-2} w_h(t') - (g_h(t') - f(r(t')) \big) \, dt' .
\end{equation}
Estimating with the aid of (\ref{tLG23q}),
\begin{equation}\label{m2avNj}
    m |\dot{r}_h(t)-\dot{r}(t)|
    \leq
    \int_0^t \big| g_h(t') - f(r(t')) \big| \, dt'
    +
    \frac{T^3}{m \kappa^2} \epsilon_h .
\end{equation}
Furthermore, by the Lipschitz continuity and sampling assumptions,
\begin{equation}
\begin{split}
    &
    \big| g_h(t) - f(r(t)) \big|
    =
    \big| f(s_h(t)) - f(r(t)) \big|
    \leq
    L
    \big| s_h(t) - r(t) \big|
    \leq \\ &
    L \,
    \big( |r_h(t) - s_h(t)| + |r_h(t) - r(t)| \big)
    \leq
    L \epsilon_h + L \, |r_h(t) - r(t)| .
\end{split}
\end{equation}
Inserting into (\ref{m2avNj}),
\begin{equation}
    m |\dot{r}_h(t)-\dot{r}(t)|
    \leq
    L \int_0^t |r_h(t') - r(t')| \, dt'
    +
    \Big( L T + \frac{T^3}{m \kappa^2} \Big) \epsilon_h .
\end{equation}
By Pointcar\'e's inequality and the first of (\ref{8qRvRa}), there is a constant $C > 0$ such that
\begin{equation}
    m |\dot{r}_h(t)-\dot{r}(t)|
    \leq
    C L \int_0^t |\dot{r}_h(t') - \dot{r}(t')| \, dt'
    +
    \Big( L T + \frac{T^3}{m\kappa^2} \Big) \epsilon_h .
\end{equation}
Finally, by the integral form of Gronwall's inequality \cite{Evans:1998},
\begin{equation}
    |\dot{r}_h(t)-\dot{r}(t)|
    \leq
    \Big[ 1 + \frac{C L}{m} t \exp\Big( \frac{C L}{m} t \Big) \Big] \,
    \Big( \frac{LT}{m} + \frac{T^3}{m^2\kappa^2} \Big) \epsilon_h ,
\end{equation}
whence we conclude that the velocity error $|\dot{r}_h(t)-\dot{r}(t)|$ and, together with the first of (\ref{8qRvRa}), the position error $|{r}_h(t)-{r}(t)|$ converge to zero uniformly over $[0,T]$, as advertised.
\end{proof}

\begin{proof}(of Prop.~\ref{dN4PE6})
Again we work directly with the Euler-Lagrange equations (\ref{8mCJjz}). Subtracting equations of motion and initial conditions, we obtain
\begin{subequations}
\begin{align}
    &
    m (\ddot{r}_h(t)-\ddot{r}(t)) + f(r_h(t),D_h) - f(r(t)) = 0 .
    \\ & \label{6Fhf5H}
    r(0) - r_h(0) = 0,
    \quad
    \dot{r}(0) - \dot{r}_h(0) = 0 .
\end{align}
\end{subequations}
Integrating with respect to time and using the second of (\ref{6Fhf5H}),
\begin{equation}
    m (\dot{r}_h(t)-\dot{r}(t))
    =
    -
    \int_0^t \big(f(r_h(t'),D_h) - f(r(t') \big) \, dt' .
\end{equation}
Estimating,
\begin{equation}\label{Ej12aT}
    m |\dot{r}_h(t)-\dot{r}(t)|
    \leq
    \int_0^t \big| f(r_h(t'),D_h) - f(r(t')) \big| \, dt' .
\end{equation}
By the Lipschitz continuity and sampling assumptions, there is $s_h \in \mathbb{R}^{3N}$ such that
\begin{equation}
\begin{split}
    &
    \big| f(r_h(t),D_h) - f(r(t)) \big|
    =
    \big| f(s_h(t)) - f(r(t)) \big|
    \leq
    L
    \big| s_h(t) - r(t) \big|
    \leq \\ &
    L \,
    \big( |r_h(t) - s_h(t)| + |r_h(t) - r(t)| \big)
    \leq
    L \epsilon_h + L \, |r_h(t) - r(t)| .
\end{split}
\end{equation}
Inserting into (\ref{Ej12aT}),
\begin{equation}
    m |\dot{r}_h(t)-\dot{r}(t)|
    \leq
    L \int_0^t |r_h(t') - r(t')| \, dt'
    +
    L T \epsilon_h.
\end{equation}
By Pointcar\'e's inequality and the first of (\ref{6Fhf5H}), there is a constant $C > 0$ such that
\begin{equation}
    m |\dot{r}_h(t)-\dot{r}(t)|
    \leq
    C L \int_0^t |\dot{r}_h(t') - \dot{r}(t')| \, dt'
    +
    L T \epsilon_h .
\end{equation}
Finally, by the integral form of Gronwall's inequality \cite{Evans:1998},
\begin{equation}
    |\dot{r}_h(t)-\dot{r}(t)|
    \leq
    \Big[ 1 + \frac{C L}{m} t \exp\Big( \frac{C L}{m} t \Big) \Big] \,
    \frac{LT}{m} \epsilon_h ,
\end{equation}
whence we conclude that the velocity error $|\dot{r}_h(t)-\dot{r}(t)|$ and, together with the first of (\ref{6Fhf5H}), the position error $|{r}_h(t)-{r}(t)|$ converge to zero uniformly over $[0,T]$, as advertised.
\end{proof}

\end{appendix}

\bibliography{biblio}
\bibliographystyle{unsrt}

\end{document}